\documentclass[a4paper,11pt]{article} 
\usepackage{jheppub}

\usepackage{tikz}
\usetikzlibrary{"arrows", "automata", "backgrounds", "calendar", "chains", "matrix", "mindmap", "patterns", "petri", "shadows", "shapes.geometric", "shapes.misc", "spy", "trees"}
\usetikzlibrary{arrows,shapes}
\usetikzlibrary{trees}
\usetikzlibrary{matrix,arrows} 	
\usetikzlibrary{positioning}				
\usetikzlibrary{calc,through}				
\usetikzlibrary{decorations.pathreplacing}
\usepackage{pgffor}

\usetikzlibrary{decorations.pathmorphing}	
\usetikzlibrary{decorations.markings}
\tikzset{
    vector/.style={decorate, decoration={snake}, draw},
	provector/.style={decorate, decoration={snake,amplitude=2.5pt}, draw},
	antivector/.style={decorate, decoration={snake,amplitude=-2.5pt}, draw},
    fermion/.style={draw=black, postaction={decorate},
        decoration={markings,mark=at position .55 with {\arrow[draw=black]{>}}}},
    fermionbar/.style={draw=black, postaction={decorate},
        decoration={markings,mark=at position .55 with {\arrow[draw=black]{<}}}},
    fermionnoarrow/.style={draw=black},
    sigma/.style={draw=green},    
    gluon/.style={decorate, draw=black,
        decoration={coil,amplitude=1.5pt, segment length=3pt}},
    phi/.style={dashed, draw=blue},
    A/.style={dotted, draw=red},
    scalar/.style={dashed,draw=black, postaction={decorate}},
    scalarbar/.style={dashed,draw=black, postaction={decorate},
        decoration={markings,mark=at position .55 with {\arrow[draw=black]{<}}}},
    scalarnoarrow/.style={dashed,draw=black},
    electron/.style={draw=black, postaction={decorate},
        decoration={markings,mark=at position .35 with {\arrow[draw=black]{>}}}},
    positron/.style={draw=black, postaction={decorate},
        decoration={markings,mark=at position .35 with {\arrow[draw=black]{<}}}},        
    bigvector/.style={decorate, decoration={snake,amplitude=4pt}, draw},
}

\usepackage[normalem]{ulem}

\newcommand{\be}{\begin{eqnarray}}
\newcommand{\ee}{\end{eqnarray}}
\newcommand{\bdm}{\begin{displaymath}}
\newcommand{\edm}{\end{displaymath}}
\newcommand{\ds}{\displaystyle}
\newcommand{\ba}{\begin{array}}
\newcommand{\ea}{\end{array}}
\newcommand{\pa}[1]{\left(#1\right)}

\allowdisplaybreaks

\newcommand{\geneva}{D\'epartement de Physique Th\'eorique and Centre for Astroparticle Physics, Universit\'e de Gen\`eve, CH-1211 Geneva, Switzerland.}
\newcommand{\natal}{International Institute of Physics, Universidade Federal do Rio Grande do Norte, Campus Universit\'ario, Lagoa Nova, Natal-RN 59078-970, Brazil.}
\newcommand{\valencia}{Instituto de F\'{\i}sica Corpuscular, Universitat de Val\`{e}ncia -- Consejo Superior de Investigaciones Cient\'{\i}ficas, Parc Cient\'{\i}fic, E-46980 Paterna, Valencia, Spain.}
\newcommand{\munich}{Max-Planck-Institut f\"ur Physik, Werner-Heisenberg-Institut, 80805 M\"unchen, Germany.}

\begin{document}

\preprint{IFIC/20-45; MPP-2020-191}

\title{\boldmath Efficient resummation of high post-Newtonian contributions to the binding energy}

\author[a]{Stefano Foffa,}
\author[b]{Riccardo Sturani,}
\author[c,d]{and William J. Torres Bobadilla}

\affiliation[a]{\geneva}
\affiliation[b]{\natal}
\affiliation[c]{\valencia}
\affiliation[d]{\munich}

\emailAdd{stefano.foffa@unige.ch}
\emailAdd{riccardo@iip.ufrn.br}
\emailAdd{torres@mpp.mpg.de}

\abstract{
A factorisation property of Feynman diagrams in the context the Effective Field Theory approach to the compact binary problem
has been recently employed to efficiently determine the static sector of the potential at fifth post-Newtonian (5PN) order. 
We extend this procedure to the case of non-static diagrams and we use it to fix, by means of elementary algebraic manipulations,
the value of more than one thousand diagrams at 5PN order, that is a substantial fraction of the diagrams needed to fully determine the dynamics at 5PN.
This procedure addresses the redundancy problem that plagues the
computation of the binding energy with respect to more ``efficient" observables
like the scattering angle, thus making the EFT approach in harmonic gauge
at least as scalable as the others methods.
}




\maketitle

\section{Introduction}
\label{sec:intro}

The determination of the binding energy of spin-less
compact binaries has recently reached the accuracy of the 4th order in the 
post-Newtonian (PN) approximation to General Relativity (GR), with several approaches obtaining the full 4PN contribution ~\cite{Damour:2014jta,Damour:2015isa,Damour:2016abl,Bernard:2015njp,Bernard:2016wrg,Bernard:2017bvn,Marchand:2017pir,Bernard:2017ktp,Foffa:2012rn,Foffa:2016rgu,Foffa:2019rdf,Foffa:2019yfl,Blumlein:2020pog}
with complete mutual agreement. The attention can now be focused on the next order, the 5PN level, 
and possibly beyond.

High-precision calculations are extremely important~\cite{Lindblom:2008cm,Antonelli:2019ytb}
  for maximising the physics output of the
  inspirals and coalescences of binary systems that are being and will be detected by the present and advanced versions of ground based 
interferometers~\cite{TheLIGOScientific:2014jea,TheVirgo:2014hva},
as well as by third generation of detectors such as ET~\cite{Punturo:2010zz} 
and by the space detector LISA~\cite{Audley:2017drz}.

Remarkably, the interplay between modern techniques originated 
in high-energy physics with classical field theory methods
has given further impulse to this quest, making
  the full determination of the 5PN sector a realistic 
goal for the near future.
More in detail, the first post-Minkowskian part (i.e. the effective potential at first order in Newton constant $G$ and at all orders in the velocity,
or 1PM for short) being known since long time~\cite{Ledvinka:2008tk}, 
it has recently been re-derived \cite{Blanchet:2018yvb,Foffa:2013gja}, 
and extended to 2PM~\cite{Damour:2017zjx}. 
Modern scattering amplitude techniques
have been employed to determine the effective potential up to 3PM and at any order in the velocity expansion~\cite{Cheung:2018wkq,Bern:2019nnu}
(not without some controversy in the interpretation of the results, see \cite{Damour:2019lcq,Bern:2020gjj,DiVecchia:2020ymx,Damour:2020tta}, confirming the necessity of multiple independent approaches to such difficult problems).

At the highest level of interaction at 5PN (that is $G^6$), the potential in the harmonic gauge has been determined using effective field theory (EFT)-based
techniques within the Non-Relativistic-General-Relativity (NRGR) approach pioneered in
\cite{Goldberger:2004jt},
in close succession by two groups in ~\cite{Foffa:2019hrb} and~\cite{Blumlein:2019zku}.
The latter by a brute force calculation,
the former exploiting a factorisation property
that applies to all static diagrams at any odd-PN level. 
The extension of this factorisation property 
to non-static diagrams is the main focus of this work.

More recently, a collaboration including the same authors as \cite{Blumlein:2019zku} has extended their results to cover some sectors of the 6PN order \cite{Blumlein:2020znm}, while the
  contribution of hereditary effects to the conservative dynamics have been determined at 5PN in \cite{Foffa:2019eeb} along the same path established for the 4PN case within the EFT approach \cite{Foffa:2011np,Porto:2017dgs}.

In addition to the above mentioned efforts using 
EFT techniques, the combination of several techniques, spanning from PN and PM expansions, to self-force approach and effective-one-body re-summation has led to a
systematisation of the center-of-mass Hamiltonian up to 6PN, 
in which the still unknown sectors have been condensed into few unknown coefficients, respectively at 5PN (in the $G^5$ and $G^6$ sector)
\cite{Bini:2019nra} and at 6PN (in the $G^5,G^6$ and $G^7$ sectors) \cite{Bini:2020nsb,Bini:2020hmy,Bini:2020uiq}.
This is a remarkable result, not quite for the small number of unknown coefficients,
but rather for the insight gained by combining information coming from the computation of observables in different limits
and within different approaches.
A similar route has been followed in \cite{Kalin:2019rwq,Kalin:2019inp,Kalin:2020fhe}, where a formalism has been developed to map
observables relative to bound states into unbound scatterings ones.

In particular, from the above analyses it has emerged that the scattering angle seems to store information about the two-body dynamics in a more efficient way than other observables. This can be seen in some remarkable properties of the scattering angle, like its simple dependence on the symmetric mass ratio \cite{Vines:2018gqi}, or the fact that it can be determined, at a given PM order, by a drastically lower number of diagrams with respect to the binding energy.

This means that, conversely, part of the information contained in the
binding energy expression at a given PM or PN order is somehow redundant, as it should be determined by the knowledge of lower order contributions.
This is exactly what has been enlightened in \cite{Foffa:2019hrb}, where it is shown that static sectors at odd-PN orders are actually ``redundant",
in the sense that they are entirely composed by Feynman diagrams
which are products of diagrams belonging to lower PN orders. 
Hence, we call such diagrams {\it factorisable}, which can be computed with simple algebraic operations, limiting the calculation of the non-factorisable ones
to the general, but computationally more demanding, method described in
\cite{Foffa:2016rgu}.

The paper is structured as follows. 
In section~\ref{sec:procedure}, we extend  the identification and computation of factorisable diagrams to the general case
(i.e. to non-static diagrams) and we provide a simple way of expressing 
factorisable diagrams in terms of lower order ones
also in presence of time derivatives.
We then apply this method to the $O(v^2)$ sector at 5PN, in section~\ref{sec:proof}, and we determine its ``redundant" contribution to the effective potential (that is, roughly two thirds of the diagrams involved in this sector) with a simple code which takes less than a minute on a normal laptop computer\footnote{Considering that factorisable diagrams represent typically at least half of the total number, 
  there is a sizable gain obtainable by a full-scale use of our method compared to
 the few hours  of CPU time mentioned in~\cite{Blumlein:2020pog} 
 or the several weeks displayed in Table 1 of~\cite{Blumlein:2020znm}.},
as it involves only elementary algebraic manipulation.
Finally, the implications of our work will be discussed 
in the conclusions section~\ref{sec:conclusion}.

\section{EFT and factorisation of non-static diagrams}
\label{sec:procedure}
The details of the procedure for computing the \emph{near-zone} (i.e.
at distances from the source smaller than the radiation wavelength)
contribution to the effective potential have been outlined and discussed
in several works~\cite{Foffa:2011ub,Foffa:2012rn,Foffa:2013qca,Foffa:2016rgu,Foffa:2019hrb,Foffa:2019rdf,Foffa:2019yfl}.
In this section, we recall just the points needed 
in this paper, starting from the fundamental path-integral equation,
\begin{equation}
\label{eq:seff}
 \exp[\text{i} S_{\rm eff}(x_1,x_2)]=\int {\cal D} g_{\mu\nu}  \exp[\text{i}S_{\rm bulk}(g_{\mu\nu})+ \text{i}S_{\rm pp}(g_{\mu\nu},x_1,x_2)]\,,
\end{equation}
with
\be\label{actions}
S_{\rm bulk}&=& 2 \Lambda^2\int {\rm d}^{d+1}x\sqrt{-g}\left[R(g)-\frac12 \Gamma^\mu\Gamma_\mu\right]\,,\\
S_{\rm pp}&=&-\!\!\sum_{a=1,2} m_a\int {\rm d}\tau_a = 
-\!\!\sum_{a=1,2} m_a\int \sqrt{-g_{\mu\nu}(x^\mu_a) {\rm d}x_a^\mu {\rm d}x_a^\nu}\,,
\ee
which relates the effective action to the fundamental actions describing Einstein gravity (plus the de-Donder gauge fixing term $-\frac12 \Gamma^\mu\Gamma_\mu$),
minimally coupled to two spin-less point particles.

Working in the post-Newtonian framework, the above expression can be perturbatively evaluated in terms of Feynman diagrams,
along the well-established formalism historically introduced in quantum field theory context and applied to the gravitationally bounded compact binary problem in \cite{Goldberger:2004jt}, also known as NRGR.
We adopt a Kaluza-Klein (KK) decomposition of the metric \cite{Blanchet:1989ki,Kol:2007bc,Kol:2007rx}
into one scalar, one spatial vector, and one symmetric tensorial field, respectively called $\phi$, $A_i$ and $\sigma_{ij}$
\be
\label{met_nr}
g_{\mu\nu}=e^{2\phi/\Lambda}\pa{
\ba{cc}
-1 & A_j/\Lambda \\
A_i/\Lambda &\quad e^{- c_d\phi/\Lambda}\gamma_{ij}-
A_iA_j/\Lambda^2\\
\ea
}\,,
\ee
with $\gamma_{ij}=\delta_{ij}+\sigma_{ij}/\Lambda$,
$c_d=2\frac{(d-1)}{(d-2)}$ and $i,j$ running over the $d$ spatial dimensions.

The fundamental bulk action in (\ref{actions}) can be expressed in terms of the KK fields, but
because of the purely algebraic method implemented in this work its explicit form is not needed here:
our method computes factorisable diagrams by relying only on the knowledge
  of previously computed, lower order diagrams and world-line vertices.

As to $S_{\rm pp}$ it reads
\be
\label{eq:ppKK}
\ba{rcl}
S_{\rm pp}&=&\ds -\sum_{a=1,2}m_a\ds \int {\rm d}\tau_a\\
&=& \ds-\sum_{a=1,2}m_a\int {\rm d}t_a\ e^{\phi/\Lambda}
\sqrt{\pa{1-\frac{A_i}{\Lambda}v^i_a}^2
-e^{-c_d \phi/\Lambda}\pa{v^2_a+\frac{\sigma_{ij}}{\Lambda}v^i_av^j_a}}\,,
\ea
\ee
where $m_a$ the mass of $a$-th particle, and it can be used to derive the particle-gravity vertex interactions with different combinations of the KK fields,
expanded up to the needed order in terms of the particle velocity $v$.
E.g. world-line vertices involving respectively $n\phi$, $n\phi$ and one $A_i$,
$n\phi$ and one $\sigma_{ij}$ fields bring the following factors from the expansion
of eq.~(\ref{eq:ppKK})
\begin{align}
\label{pver}
&&\parbox{20mm}{
\begin{tikzpicture}[line width=1 pt,node distance=0.4 cm and 0.4 cm]
\coordinate[] (v1);
\coordinate[right = of v1] (v2);
\coordinate[right = of v2] (v3);
\coordinate[right = of v3] (v4);
\coordinate[right = of v4] (v5);
\coordinate[below = of v3] (v9);
\coordinate[below = of v9, label=center: $\overset{\cdots}{n}$] (v6);
\coordinate[left = of v6] (v7);
\coordinate[right = of v6] (v8);
\draw[phi] (v3) -- (v7);
\draw[phi] (v3) -- (v8);
\draw[fermionnoarrow] (v1) -- (v5);
\end{tikzpicture}
}
\!\!\!
\simeq&
-\frac{\text{i}\ m}{n!  \Lambda^n}\left[1-\frac{d^n}{2(2-d)^n} v^2\right]\,,
\notag\\
&&\parbox{20mm}{
\begin{tikzpicture}[line width=1 pt,node distance=0.4 cm and 0.4 cm]
\coordinate[] (v1);
\coordinate[right = of v1] (v2);
\coordinate[right = of v2] (v3);
\coordinate[right = of v3] (v4);
\coordinate[right = of v4] (v5);
\coordinate[below = of v3] (v9);
\coordinate[below = of v9, label=center: $\overset{\cdots}{n}$] (v6);
\coordinate[left = of v6] (v7);
\coordinate[left = of v7] (v10);
\coordinate[right = of v6] (v8);
\draw[A] (v3) -- (v10);
\draw[phi] (v3) -- (v7);
\draw[phi] (v3) -- (v8);
\draw[fermionnoarrow] (v1) -- (v5);
\end{tikzpicture}
}
\!\!\!
\simeq&
\frac{\text{i}\ m v^i}{n!  \Lambda^{n+1}}\,,
&&
\parbox{20mm}{
\begin{tikzpicture}[line width=1 pt,node distance=0.4 cm and 0.4 cm]
\coordinate[] (v1);
\coordinate[right = of v1] (v2);
\coordinate[right = of v2] (v3);
\coordinate[right = of v3] (v4);
\coordinate[right = of v4] (v5);
\coordinate[below = of v3] (v9);
\coordinate[below = of v9, label=center: $\overset{\cdots}{n}$] (v6);
\coordinate[left = of v6] (v7);
\coordinate[left = of v7] (v10);
\coordinate[right = of v6] (v8);
\draw[sigma] (v3) -- (v10);
\draw[phi] (v3) -- (v7);
\draw[phi] (v3) -- (v8);
\draw[fermionnoarrow] (v1) -- (v5);
\end{tikzpicture}
}
\!\!\!
\simeq&
\frac{\text{i}\ m v^i v^j}{n!  \Lambda^{n+1}}\left[\frac{d}{2(2-d)}\right]^n\,.
\end{align}
Black, dashed blue, dotted red, and green lines stand respectively for the particle world-line and the 
$\phi$, $A_i$ and $\sigma_{ij}$ propagators, while $\Lambda^{-2}\equiv 32 \pi G L^{d-3}$ is the $d$-dimensional gravitational coupling
(we work in dimensional regularisation, with the arbitrary length-scale $L$ eventually dropping out from observables).

To elucidate the notion of factorisable diagrams, let us consider a static 5PN
diagram, together with its $p$-Fourier transformation, being $p$ the
three-momentum exchanged between the two massive particles.
In general, any gravity-amplitude of order $G_N^\ell$ can be
mapped onto an $(\ell-1)$-loop 2-point function with massless internal
lines and external momentum $p$. 
Then, after taking the $p$-Fourier transformation, one is led e.g. to,
\begin{equation}
\int\frac{d^{\rm d}p}{\pa{2\pi}^{\rm d}} \ e^{\text{i} p \cdot r} \
\parbox{15mm}{
\begin{tikzpicture}[line width=1 pt,node distance= 0.6 cm and 0.3 cm]
\coordinate[] (v0);
\coordinate[right = of v0] (v1);
\coordinate[right = of v1] (v2);
\coordinate[right = of v2] (v2a);
\coordinate[right = of v2a] (v6);
\coordinate[right = of v6] (v7);
\coordinate[above right = of v7] (v9);
\coordinate[below right = of v9] (v8);
\coordinate[right = of v8] (v12);
\coordinate[above = of v2] (v3);
\coordinate[above = of v3] (v4);
\coordinate[left = of v4] (v4a);
\coordinate[left = of v4a] (v11);
\coordinate[right = of v4] (v4b);
\coordinate[right = of v4b] (v5);
\coordinate[above right = of v9] (v9a);
\coordinate[right = of v9a] (v10);
\draw[phi] (v7) -- (v9);
\draw[phi] (v8) -- (v9);
\draw[phi] (v5) -- (v6);
\draw[phi] (v4) -- (v2);
\draw[phi] (v4) -- (v1);
\draw[sigma] (v3) -- (v9);
\draw[fermionnoarrow] (v0) -- (v12);
\draw[fermionnoarrow] (v11) -- (v10);
\end{tikzpicture}
}
\qquad\quad\to\quad
\parbox{15mm}{
\begin{tikzpicture}[line width=1 pt,node distance= 0.3 cm and 0.3 cm]
\coordinate[] (v0);
\coordinate[above = of v0] (v1);
\coordinate[below = of v0] (v2);
\coordinate[left = of v0] (v3a);
\coordinate[left = of v3a] (v3);
\coordinate[right = of v0] (v4a);
\coordinate[right = of v4a] (v4);
\draw[blue,dashed] (v0) circle (.6cm);
\draw[sigma] (v1) -- (v3);
\draw[sigma] (v2) -- (v3);
\draw[phi] (v4) arc (30:100:0.6cm);
\draw[phi] (v4) arc (-30:-100:0.6cm);
\draw[phi] (v1) arc (210:280:0.6cm);
\draw[phi] (v2) arc (-210:-280:0.6cm);
\draw[fill=black] (v4) circle (.05cm);
\draw[phi] (v4) arc (180:-180:0.225cm);
\end{tikzpicture}
}
\quad\ ,
\end{equation}
where in the final representation the external legs of the 1-point function
have been joined and collapsed to a point to indicate the residual $r$
dependence of the diagram, finally represented by product of massless
vacuum-vacuum diagrams.

As in~\cite{Foffa:2019hrb}, it has been observed that the value of
the diagram on the left can be computed in terms of the values of the two
sub-diagrams,
in which it is naturally subdivided (above, on the right), according to a simple formula,
\begin{eqnarray}\label{prod}
{\cal V}^{\rm factorisable} = \Big({\cal V}_1 \times {\cal
  V}_{2}\Big) \times {\cal K}\times{\cal C}\,,
\end{eqnarray}
with,
{\it i)} ${\cal V}$ and ${\cal V}_{1,2}$ are, respectively, the values of the factorisable diagram and of the two sub-diagrams,
{\it ii)} ${\cal K}$ accounts for the new matter interaction vertex of ${\cal V}$ (emerging from the sewing) out of matter interaction vertices
of the two sub-diagrams and can be determined
using equation (\ref{pver}), and {\it iii)} ${\cal C}=C^{\rm factorisable}/(C_{1} \times C_{2})$ where the $C$'s are the combinatoric factors associated with each graph.

This procedure, introduced in~\cite{Foffa:2019hrb}, where it
was used to obtain all the contributions to the static part of the $5$PN effective action,
can be extended to the non-static case with only minor adjustments, the main difference being that non-static diagrams can contain time derivatives which can ``propagate'' across factors
 a feature that can be naturally implemented in the
codes usually employed to write the amplitude corresponding to a given graph.

\begin{figure}[h]
\begin{center}
\includegraphics[width=.25\linewidth]{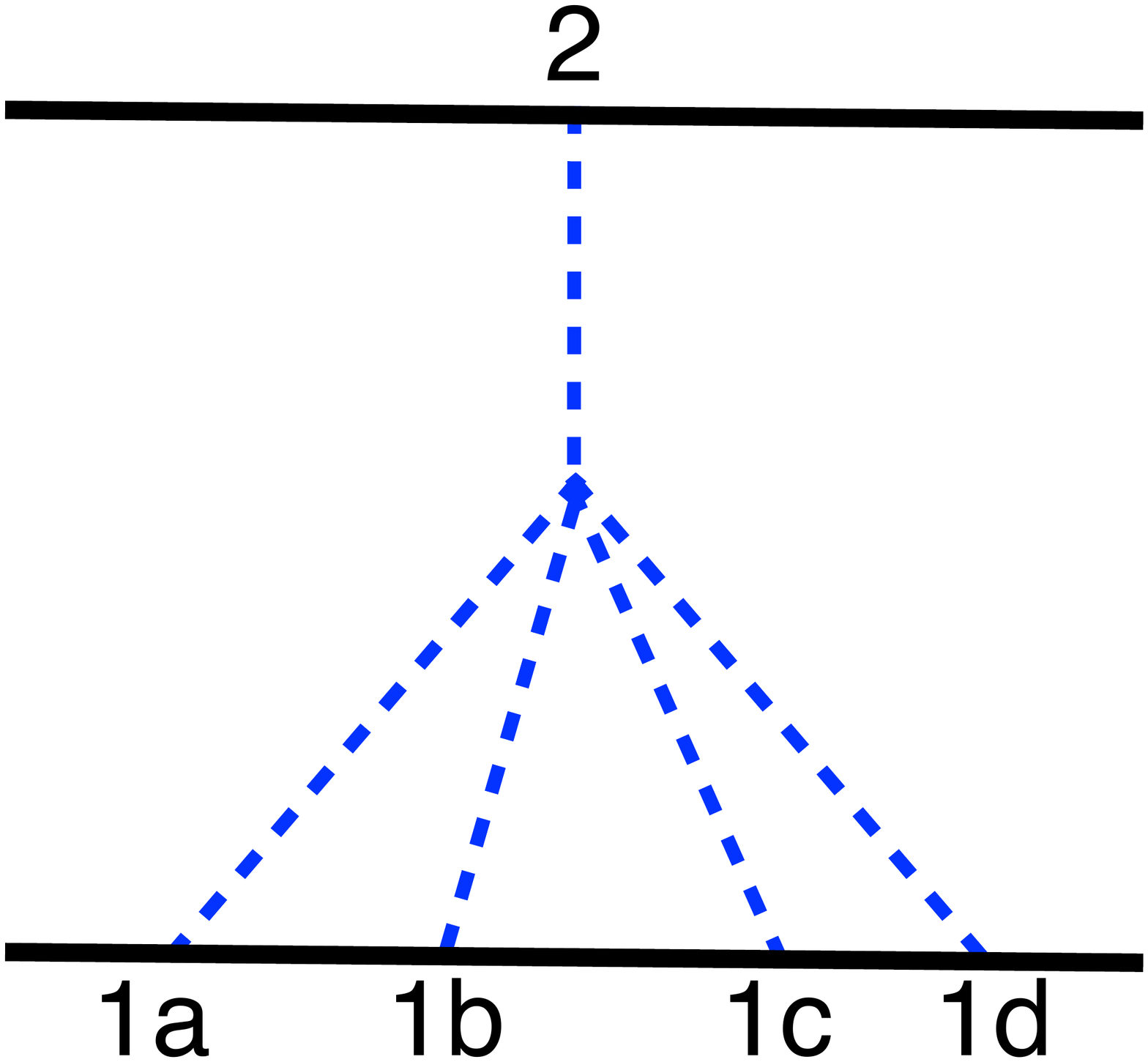}
\includegraphics[width=.25\linewidth]{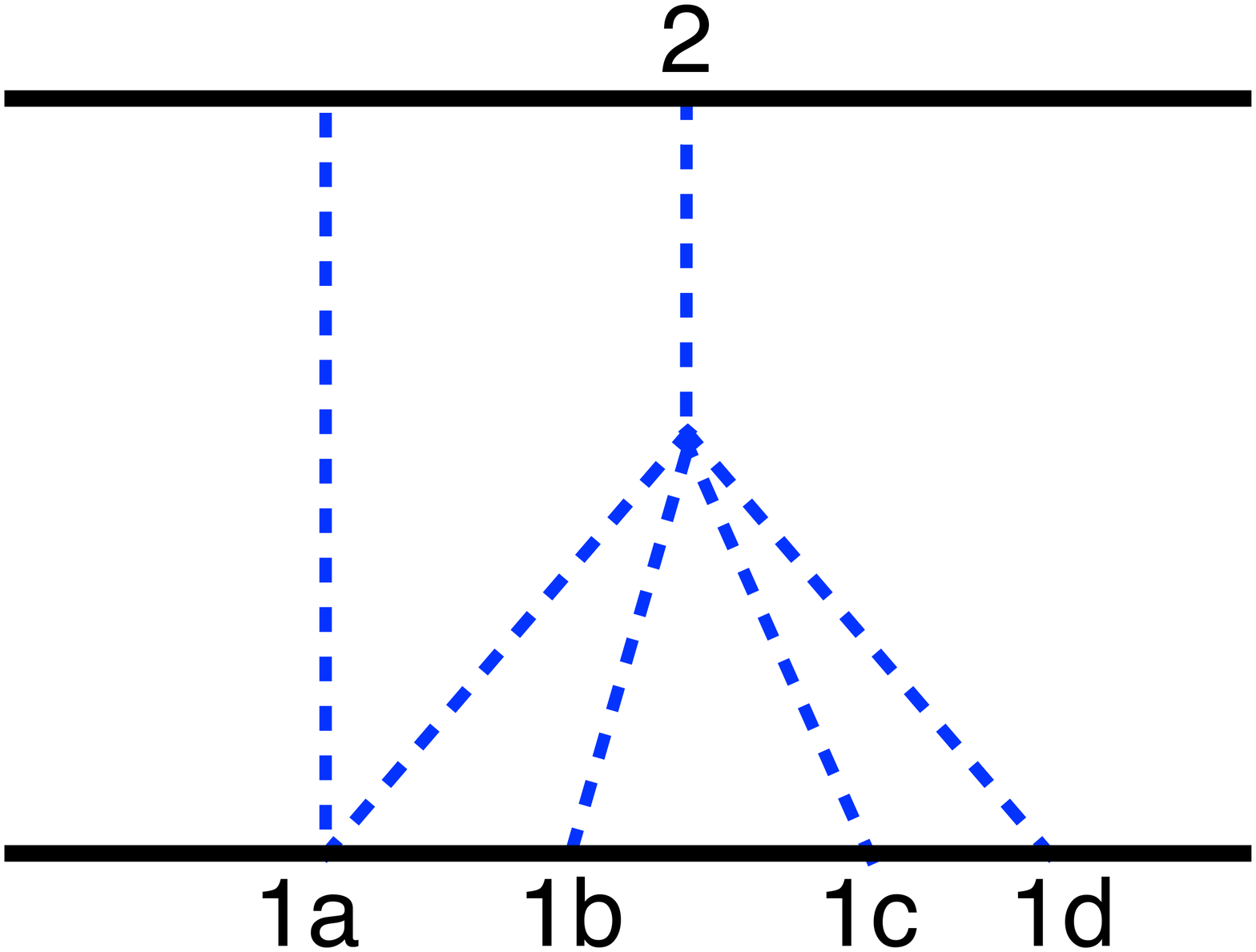}
\caption{A 3-loop non-factorisable diagram (left) and a 4-loop factorisable one (right) which can be computed as the product of the former with the Newtonian one.}
\label{fig:prime}
\end{center}
\end{figure}

For instance, the 3-loop diagram in the left of figure~\ref{fig:prime}
contains two times derivatives in the bulk $\phi^5$ vertex and has the following value,
\begin{align}
\label{Deq}
%
\frac{G^{4}m_{1}^{4}m_{2}}{r^{4}}\Bigg[&\frac{8}{3}v_{1}.v_{2}-\frac{4}{9}v_{1}^{2}+\frac{8}{3}{v_{1}^{r}}^{2}-\frac{16}{9}v_{1}^{r}v_{2}^{r}\notag\\
&+\frac{2}{3}\left(4v_{2}^{r}-3v_{1}^{r}\right)\left(D_{1a}+D_{1b}+D_{1c}+D_{1d}\right)-\frac{8}{3}v_{1}^{r}D_{2}\notag\\
&+\frac{2}{3}\left(2D_{2}+D_{1a}+D_{1b}+D_{1c}+D_{1d}\right)\left(D_{1a}+D_{1b}+D_{1c}+D_{1d}\right)\notag\\
&-\frac{2}{3}\left(D_{1a}^{2}+D_{1b}^{2}+D_{1c}^{2}+D_{1d}^{2}\right)\Bigg]\,.
\end{align}
The symbols $D_{ai}$ represent time derivative operators acting
  on particle $a=1,2$ at world-line vertex $i$ on
  whatever sub-graph is eventually attached there (the index $i$ has been
  omitted when particle $a$ has only one vertex).
By setting the $D_{ai}$'s 
to zero, one obtains the value of the diagram which contributes to the 
4PN effective action as computed in~\cite{Foffa:2019rdf}.

In this way, equation~(\ref{prod}) holds also in the non-static case, provided the $\times$ symbol between ${\cal V}_1$ and ${\cal V}_2$ is interpreted in an extensive way to include also the action of the derivative operators $D_{ai}$'s.  For instance, if one wants to compute the value of the 4-loop diagram on the right part of figure \ref{fig:prime}, one has simply to ``multiply" it by the value of the Newtonian diagram $N\equiv-\frac{G m_1 m_2}r$, setting all the $D_i$'s to zero except for $D_{1a}$, which instead turns $N$ into $\dot{N}$. 
After inserting the appropriate factors ${\cal K}$ and ${\cal C}$ (see the appendix~\ref{appendix} for more details) one obtains the result,
\be
\frac89\frac{G^5 m_1^4 m_2^2}{r^5}\left[2v_1^2 -12 v_1.v_2 -21 {v_1^r}^2+69v_1^r v_2^r-12{v_2^r}^2\right]\,,
\ee
without the need of performing any 4-loop momentum integration.
A more involved example is discussed in detail in appendix~\ref{appendix}.

\section{Proof of concept}
\label{sec:proof}
In this section, we apply the method described above to the determination of factorisable diagrams up to order $v^2$ and up to 5PN
(this procedure is generalisable to higher powers of $v$ and we leave this to future work).
As a first step, we have modified our codes to include derivative operators, as in eq.~(\ref{Deq}), and we have recomputed all the non-factorisable Feynman diagrams containing up to two derivatives, and up to 4PN, to take the $D_i$
operators into account.
Second, we have used these prime diagrams as building blocks to generate all the factorisable ones.
We have verified that we have been able in this way to reproduce the value of all the factorisable diagrams up to 4PN, recovering the values already computed in \cite{Gilmore:2008gq,Foffa:2011ub,Foffa:2019rdf}.

Next, we moved on to 5PN, and we provide here a schematic summary of the result, where the 1220 $G^5v^2$ factorisable diagrams (roughly two thirds of the total) have been divided into 8 subcategories.
\noindent
The diagram with a single $A_i$ propagator can be combined with four Newtonian diagrams, as schematically represented below
\begin{align}
\left(\,\parbox{15mm}{
\begin{tikzpicture}[line width=1 pt,node distance=0.5 cm and 0.3125 cm]
\coordinate[] (v1);
\coordinate[right = of v1] (v2);
\coordinate[right = of v2] (v3);
\coordinate[below = of v2] (v4);
\coordinate[below = of v4] (v5);
\coordinate[left = of v5] (v6);
\coordinate[right = of v5] (v7);
\draw[fermionnoarrow] (v1) -- (v3);
\draw[A] (v2) -- (v5);
\draw[fermionnoarrow] (v6) -- (v7);
\end{tikzpicture}
}\!\!\!\!\!\!\!\!\!\!\!\!\right)
&\times
\left(\,\parbox{15mm}{
\begin{tikzpicture}[line width=1 pt,node distance=0.5 cm and 0.3125 cm]
\coordinate[] (v1);
\coordinate[right = of v1] (v2);
\coordinate[right = of v2] (v3);
\coordinate[below = of v2] (v4);
\coordinate[below = of v4] (v5);
\coordinate[left = of v5] (v6);
\coordinate[right = of v5] (v7);
\draw[fermionnoarrow] (v1) -- (v3);
\draw[phi] (v2) -- (v5);
\draw[fermionnoarrow] (v6) -- (v7);
\end{tikzpicture}
}\!\!\!\!\!\!\!\!\!\!\!\!\right)^4\\
&=\left\{\ \parbox{15mm}{
\begin{tikzpicture}[line width=1 pt,node distance=0.5 cm and 0.3125 cm]
\coordinate[] (v1);
\coordinate[left = of v1] (v0);
\coordinate[right = of v1] (v2);
\coordinate[right = of v2] (v3);
\coordinate[right = of v3] (v8);
\coordinate[below = of v2] (v4);
\coordinate[below = of v4] (v5);
\coordinate[left = of v5] (v6);
\coordinate[left = of v6] (v9);
\coordinate[left = of v6] (v11);
\coordinate[right = of v5] (v7);
\coordinate[right = of v7] (v10);
\draw[fermionnoarrow] (v0) -- (v8);
\draw[phi] (v2) -- (v9);
\draw[phi] (v2) -- (v5);
\draw[phi] (v2) -- (v6);
\draw[phi] (v2) -- (v7);
\draw[A] (v2) -- (v10);
\draw[fermionnoarrow] (v11) -- (v10);
\end{tikzpicture}
}
 \parbox{15mm}{
\begin{tikzpicture}[line width=1 pt,node distance=0.5 cm and 0.3125 cm]
\coordinate[] (v1);
\coordinate[left = of v1] (v0);
\coordinate[right = of v1] (v2);
\coordinate[right = of v2] (v3);
\coordinate[right = of v3] (v8);
\coordinate[below = of v2] (v4);
\coordinate[below = of v4] (v5);
\coordinate[left = of v5] (v6);
\coordinate[left = of v6] (v9);
\coordinate[left = of v6] (v11);
\coordinate[right = of v5] (v7);
\coordinate[right = of v7] (v10);
\draw[fermionnoarrow] (v0) -- (v8);
\draw[phi] (v2) -- (v9);
\draw[phi] (v2) -- (v5);
\draw[phi] (v2) -- (v6);
\draw[phi] (v2) -- (v7);
\draw[A] (v7) -- (v3);
\draw[fermionnoarrow] (v11) -- (v10);
\end{tikzpicture}
}
\parbox{15mm}{
\begin{tikzpicture}[line width=1 pt,node distance=0.5 cm and 0.3125 cm]
\coordinate[] (v1);
\coordinate[left = of v1] (v0);
\coordinate[right = of v1] (v2);
\coordinate[right = of v2] (v3);
\coordinate[right = of v3] (v8);
\coordinate[below = of v2] (v4);
\coordinate[below = of v4] (v5);
\coordinate[left = of v5] (v6);
\coordinate[left = of v6] (v9);
\coordinate[left = of v6] (v11);
\coordinate[right = of v5] (v7);
\coordinate[right = of v7] (v10);
\draw[fermionnoarrow] (v0) -- (v8);
\draw[phi] (v2) -- (v9);
\draw[phi] (v2) -- (v5);
\draw[phi] (v2) -- (v6);
\draw[A] (v2) -- (v7);
\draw[phi] (v7) -- (v3);
\draw[fermionnoarrow] (v11) -- (v10);
\end{tikzpicture}
}
\parbox{15mm}{
\begin{tikzpicture}[line width=1 pt,node distance=0.5 cm and 0.3125 cm]
\coordinate[] (v1);
\coordinate[left = of v1] (v0);
\coordinate[right = of v1] (v2);
\coordinate[right = of v2] (v3);
\coordinate[right = of v3] (v8);
\coordinate[below = of v2] (v4);
\coordinate[below = of v4] (v5);
\coordinate[left = of v5] (v6);
\coordinate[left = of v6] (v9);
\coordinate[left = of v6] (v11);
\coordinate[right = of v5] (v7);
\coordinate[right = of v7] (v10);
\draw[fermionnoarrow] (v0) -- (v8);
\draw[phi] (v2) -- (v6);
\draw[phi] (v2) -- (v5);
\draw[phi] (v2) -- (v7);
\draw[phi] (v7) -- (v3);
\draw[A] (v2) -- (v9);
\draw[fermionnoarrow] (v11) -- (v10);
\end{tikzpicture}
}
\parbox{15mm}{
\begin{tikzpicture}[line width=1 pt,node distance=0.5 cm and 0.3125 cm]
\coordinate[] (v1);
\coordinate[left = of v1] (v0);
\coordinate[right = of v1] (v2);
\coordinate[right = of v2] (v3);
\coordinate[right = of v3] (v8);
\coordinate[below = of v2] (v4);
\coordinate[below = of v4] (v5);
\coordinate[left = of v5] (v6);
\coordinate[left = of v6] (v9);
\coordinate[left = of v6] (v11);
\coordinate[right = of v5] (v7);
\coordinate[right = of v7] (v10);
\draw[fermionnoarrow] (v0) -- (v8);
\draw[phi] (v2) -- (v9);
\draw[phi] (v2) -- (v5);
\draw[phi] (v2) -- (v6);
\draw[phi] (v5) -- (v3);
\draw[A] (v3) -- (v7);
\draw[fermionnoarrow] (v11) -- (v10);
\end{tikzpicture}
}
\parbox{15mm}{
\begin{tikzpicture}[line width=1 pt,node distance=0.5 cm and 0.3125 cm]
\coordinate[] (v1);
\coordinate[left = of v1] (v0);
\coordinate[right = of v1] (v2);
\coordinate[right = of v2] (v3);
\coordinate[right = of v3] (v8);
\coordinate[below = of v2] (v4);
\coordinate[below = of v4] (v5);
\coordinate[left = of v5] (v6);
\coordinate[left = of v6] (v9);
\coordinate[left = of v6] (v11);
\coordinate[right = of v5] (v7);
\coordinate[right = of v7] (v10);
\draw[fermionnoarrow] (v0) -- (v8);
\draw[phi] (v2) -- (v9);
\draw[phi] (v2) -- (v5);
\draw[phi] (v2) -- (v6);
\draw[A] (v5) -- (v3);
\draw[phi] (v3) -- (v7);
\draw[fermionnoarrow] (v11) -- (v10);
\end{tikzpicture}
}
\parbox{15mm}{
\begin{tikzpicture}[line width=1 pt,node distance=0.5 cm and 0.3125 cm]
\coordinate[] (v1);
\coordinate[left = of v1] (v0);
\coordinate[right = of v1] (v2);
\coordinate[right = of v2] (v3);
\coordinate[right = of v3] (v8);
\coordinate[below = of v2] (v4);
\coordinate[below = of v4] (v5);
\coordinate[left = of v5] (v6);
\coordinate[left = of v6] (v9);
\coordinate[left = of v6] (v11);
\coordinate[right = of v5] (v7);
\coordinate[right = of v7] (v10);
\draw[fermionnoarrow] (v0) -- (v8);
\draw[phi] (v2) -- (v9);
\draw[A] (v2) -- (v5);
\draw[phi] (v2) -- (v6);
\draw[phi] (v5) -- (v3);
\draw[phi] (v3) -- (v7);
\draw[fermionnoarrow] (v11) -- (v10);
\end{tikzpicture}
}
\parbox{15mm}{
\begin{tikzpicture}[line width=1 pt,node distance=0.5 cm and 0.3125 cm]
\coordinate[] (v1);
\coordinate[left = of v1] (v0);
\coordinate[right = of v1] (v2);
\coordinate[right = of v2] (v3);
\coordinate[right = of v3] (v8);
\coordinate[below = of v2] (v4);
\coordinate[below = of v4] (v5);
\coordinate[left = of v5] (v6);
\coordinate[left = of v6] (v9);
\coordinate[left = of v6] (v11);
\coordinate[right = of v5] (v7);
\coordinate[right = of v7] (v10);
\draw[fermionnoarrow] (v0) -- (v8);
\draw[A] (v2) -- (v9);
\draw[phi] (v2) -- (v5);
\draw[phi] (v2) -- (v6);
\draw[phi] (v5) -- (v3);
\draw[phi] (v3) -- (v7);
\draw[fermionnoarrow] (v11) -- (v10);
\end{tikzpicture}
}
\right.\nonumber\\
&\quad
\left.\parbox{15mm}{
\begin{tikzpicture}[line width=1 pt,node distance=0.5 cm and 0.3125 cm]
\coordinate[] (v1);
\coordinate[left = of v1] (v0);
\coordinate[right = of v1] (v2);
\coordinate[right = of v2] (v3);
\coordinate[right = of v3] (v8);
\coordinate[below = of v2] (v4);
\coordinate[below = of v4] (v5);
\coordinate[left = of v5] (v6);
\coordinate[left = of v6] (v9);
\coordinate[left = of v6] (v11);
\coordinate[right = of v5] (v7);
\coordinate[right = of v7] (v10);
\draw[fermionnoarrow] (v0) -- (v8);
\draw[phi] (v1) -- (v9);
\draw[phi] (v1) -- (v5);
\draw[phi] (v1) -- (v6);
\draw[phi] (v5) -- (v2);
\draw[A] (v5) -- (v3);
\draw[fermionnoarrow] (v11) -- (v10);
\end{tikzpicture}
}
\parbox{15mm}{
\begin{tikzpicture}[line width=1 pt,node distance=0.5 cm and 0.3125 cm]
\coordinate[] (v1);
\coordinate[left = of v1] (v0);
\coordinate[right = of v1] (v2);
\coordinate[right = of v2] (v3);
\coordinate[right = of v3] (v8);
\coordinate[below = of v2] (v4);
\coordinate[below = of v4] (v5);
\coordinate[left = of v5] (v6);
\coordinate[left = of v6] (v9);
\coordinate[left = of v6] (v11);
\coordinate[right = of v5] (v7);
\coordinate[right = of v7] (v10);
\draw[fermionnoarrow] (v0) -- (v8);
\draw[phi] (v1) -- (v9);
\draw[A] (v1) -- (v5);
\draw[phi] (v1) -- (v6);
\draw[phi] (v5) -- (v2);
\draw[phi] (v5) -- (v3);
\draw[fermionnoarrow] (v11) -- (v10);
\end{tikzpicture}
}
\parbox{15mm}{
\begin{tikzpicture}[line width=1 pt,node distance=0.5 cm and 0.3125 cm]
\coordinate[] (v1);
\coordinate[left = of v1] (v0);
\coordinate[right = of v1] (v2);
\coordinate[right = of v2] (v3);
\coordinate[right = of v3] (v8);
\coordinate[below = of v2] (v4);
\coordinate[below = of v4] (v5);
\coordinate[left = of v5] (v6);
\coordinate[left = of v6] (v9);
\coordinate[left = of v6] (v11);
\coordinate[right = of v5] (v7);
\coordinate[right = of v7] (v10);
\draw[fermionnoarrow] (v0) -- (v8);
\draw[phi] (v2) -- (v6);
\draw[phi] (v2) -- (v5);
\draw[phi] (v2) -- (v7);
\draw[A] (v7) -- (v3);
\draw[phi] (v6) -- (v1);
\draw[fermionnoarrow] (v11) -- (v10);
\end{tikzpicture}
}
\parbox{15mm}{
\begin{tikzpicture}[line width=1 pt,node distance=0.5 cm and 0.3125 cm]
\coordinate[] (v1);
\coordinate[left = of v1] (v0);
\coordinate[right = of v1] (v2);
\coordinate[right = of v2] (v3);
\coordinate[right = of v3] (v8);
\coordinate[below = of v2] (v4);
\coordinate[below = of v4] (v5);
\coordinate[left = of v5] (v6);
\coordinate[left = of v6] (v9);
\coordinate[left = of v6] (v11);
\coordinate[right = of v5] (v7);
\coordinate[right = of v7] (v10);
\draw[fermionnoarrow] (v0) -- (v8);
\draw[phi] (v2) -- (v6);
\draw[phi] (v2) -- (v5);
\draw[A] (v2) -- (v7);
\draw[phi] (v7) -- (v3);
\draw[phi] (v6) -- (v1);
\draw[fermionnoarrow] (v11) -- (v10);
\end{tikzpicture}
}
\parbox{15mm}{
\begin{tikzpicture}[line width=1 pt,node distance=0.5 cm and 0.3125 cm]
\coordinate[] (v1);
\coordinate[left = of v1] (v0);
\coordinate[right = of v1] (v2);
\coordinate[right = of v2] (v3);
\coordinate[right = of v3] (v8);
\coordinate[below = of v2] (v4);
\coordinate[below = of v4] (v5);
\coordinate[left = of v5] (v6);
\coordinate[left = of v6] (v9);
\coordinate[left = of v6] (v11);
\coordinate[right = of v5] (v7);
\coordinate[right = of v7] (v10);
\draw[fermionnoarrow] (v0) -- (v8);
\draw[phi] (v2) -- (v6);
\draw[A] (v2) -- (v5);
\draw[phi] (v2) -- (v7);
\draw[phi] (v7) -- (v3);
\draw[phi] (v6) -- (v1);
\draw[fermionnoarrow] (v11) -- (v10);
\end{tikzpicture}
}
\parbox{15mm}{
\begin{tikzpicture}[line width=1 pt,node distance=0.5 cm and 0.3125 cm]
\coordinate[] (v1);
\coordinate[left = of v1] (v0);
\coordinate[right = of v1] (v2);
\coordinate[right = of v2] (v3);
\coordinate[right = of v3] (v8);
\coordinate[below = of v2] (v4);
\coordinate[below = of v4] (v5);
\coordinate[left = of v5] (v6);
\coordinate[left = of v6] (v9);
\coordinate[left = of v6] (v11);
\coordinate[right = of v5] (v7);
\coordinate[right = of v7] (v10);
\draw[fermionnoarrow] (v0) -- (v8);
\draw[phi] (v1) -- (v6);
\draw[phi] (v6) -- (v2);
\draw[phi] (v2) -- (v5);
\draw[phi] (v5) -- (v3);
\draw[A] (v3) -- (v7);
\draw[fermionnoarrow] (v11) -- (v10);
\end{tikzpicture}
}
\parbox{15mm}{
\begin{tikzpicture}[line width=1 pt,node distance=0.5 cm and 0.3125 cm]
\coordinate[] (v1);
\coordinate[left = of v1] (v0);
\coordinate[right = of v1] (v2);
\coordinate[right = of v2] (v3);
\coordinate[right = of v3] (v8);
\coordinate[below = of v2] (v4);
\coordinate[below = of v4] (v5);
\coordinate[left = of v5] (v6);
\coordinate[left = of v6] (v9);
\coordinate[left = of v6] (v11);
\coordinate[right = of v5] (v7);
\coordinate[right = of v7] (v10);
\draw[fermionnoarrow] (v0) -- (v8);
\draw[phi] (v1) -- (v6);
\draw[phi] (v6) -- (v2);
\draw[phi] (v2) -- (v5);
\draw[A] (v5) -- (v3);
\draw[phi] (v3) -- (v7);
\draw[fermionnoarrow] (v11) -- (v10);
\end{tikzpicture}
}
\parbox{15mm}{
\begin{tikzpicture}[line width=1 pt,node distance=0.5 cm and 0.3125 cm]
\coordinate[] (v1);
\coordinate[left = of v1] (v0);
\coordinate[right = of v1] (v2);
\coordinate[right = of v2] (v3);
\coordinate[right = of v3] (v8);
\coordinate[below = of v2] (v4);
\coordinate[below = of v4] (v5);
\coordinate[left = of v5] (v6);
\coordinate[left = of v6] (v9);
\coordinate[left = of v6] (v11);
\coordinate[right = of v5] (v7);
\coordinate[right = of v7] (v10);
\draw[fermionnoarrow] (v0) -- (v8);
\draw[phi] (v1) -- (v6);
\draw[phi] (v6) -- (v2);
\draw[A] (v2) -- (v5);
\draw[phi] (v5) -- (v3);
\draw[phi] (v3) -- (v7);
\draw[fermionnoarrow] (v11) -- (v10);
\end{tikzpicture}
}\right\}\nonumber
\end{align}
to give 16 $G^5v^2$ factorisable diagrams, whose total contribution to the potential is
\begin{align}
V_{{N}^4Gv^2}&=\frac16\frac{G^5 m_1^5m_2}{r^5}v_1.v_2+\frac{40}3\frac{G^5 m_1^4m_2^2}{r^5}v_1.v_2+\frac{45}2\frac{G^5 m_1^3m_2^3}{r^5}v_1.v_2 + 
(1\!\leftrightarrow\!2)\,.
\end{align}
\noindent
The combination of three Newtonian diagrams with the 7 $G^2v^2$ prime diagrams
\begin{eqnarray}
&&
\left(\ \parbox{15mm}{
\begin{tikzpicture}[line width=1 pt,node distance=0.5 cm and 0.3125 cm]
\coordinate[] (v1);
\coordinate[left = of v1] (v0);
\coordinate[right = of v1] (v2);
\coordinate[right = of v2] (v3);
\coordinate[right = of v3] (v8);
\coordinate[below = of v2] (v4);
\coordinate[below = of v4] (v5);
\coordinate[left = of v5] (v6);
\coordinate[left = of v6] (v9);
\coordinate[right = of v5] (v7);
\coordinate[right = of v7] (v10);
\draw[fermionnoarrow] (v0) -- (v8);
\draw[phi] (v2) -- (v4);
\draw[phi] (v4) -- (v6);
\draw[phi] (v4) -- (v7);
\draw[fermionnoarrow] (v9) -- (v10);
\end{tikzpicture}
}
\parbox{15mm}{
\begin{tikzpicture}[line width=1 pt,node distance=0.5 cm and 0.3125 cm]
\coordinate[] (v1);
\coordinate[left = of v1] (v0);
\coordinate[right = of v1] (v2);
\coordinate[right = of v2] (v3);
\coordinate[right = of v3] (v8);
\coordinate[below = of v2] (v4);
\coordinate[below = of v4] (v5);
\coordinate[left = of v5] (v6);
\coordinate[left = of v6] (v9);
\coordinate[right = of v5] (v7);
\coordinate[right = of v7] (v10);
\draw[fermionnoarrow] (v0) -- (v8);
\draw[phi] (v2) -- (v4);
\draw[A] (v4) -- (v6);
\draw[phi] (v4) -- (v7);
\draw[fermionnoarrow] (v9) -- (v10);
\end{tikzpicture}
}
\parbox{15mm}{
\begin{tikzpicture}[line width=1 pt,node distance=0.5 cm and 0.3125 cm]
\coordinate[] (v1);
\coordinate[left = of v1] (v0);
\coordinate[right = of v1] (v2);
\coordinate[right = of v2] (v3);
\coordinate[right = of v3] (v8);
\coordinate[below = of v2] (v4);
\coordinate[below = of v4] (v5);
\coordinate[left = of v5] (v6);
\coordinate[left = of v6] (v9);
\coordinate[right = of v5] (v7);
\coordinate[right = of v7] (v10);
\draw[fermionnoarrow] (v0) -- (v8);
\draw[A] (v2) -- (v4);
\draw[phi] (v4) -- (v6);
\draw[phi] (v4) -- (v7);
\draw[fermionnoarrow] (v9) -- (v10);
\end{tikzpicture}
}
\parbox{15mm}{
\begin{tikzpicture}[line width=1 pt,node distance=0.5 cm and 0.3125 cm]
\coordinate[] (v1);
\coordinate[left = of v1] (v0);
\coordinate[right = of v1] (v2);
\coordinate[right = of v2] (v3);
\coordinate[right = of v3] (v8);
\coordinate[below = of v2] (v4);
\coordinate[below = of v4] (v5);
\coordinate[left = of v5] (v6);
\coordinate[left = of v6] (v9);
\coordinate[right = of v5] (v7);
\coordinate[right = of v7] (v10);
\draw[fermionnoarrow] (v0) -- (v8);
\draw[A] (v2) -- (v4);
\draw[A] (v4) -- (v6);
\draw[phi] (v4) -- (v7);
\draw[fermionnoarrow] (v9) -- (v10);
\end{tikzpicture}
}
\parbox{15mm}{
\begin{tikzpicture}[line width=1 pt,node distance=0.5 cm and 0.3125 cm]
\coordinate[] (v1);
\coordinate[left = of v1] (v0);
\coordinate[right = of v1] (v2);
\coordinate[right = of v2] (v3);
\coordinate[right = of v3] (v8);
\coordinate[below = of v2] (v4);
\coordinate[below = of v4] (v5);
\coordinate[left = of v5] (v6);
\coordinate[left = of v6] (v9);
\coordinate[right = of v5] (v7);
\coordinate[right = of v7] (v10);
\draw[fermionnoarrow] (v0) -- (v8);\draw[phi] (v2) -- (v4);
\draw[A] (v4) -- (v6);
\draw[A] (v4) -- (v7);
\draw[fermionnoarrow] (v9) -- (v10);
\end{tikzpicture}
}
\parbox{15mm}{
\begin{tikzpicture}[line width=1 pt,node distance=0.5 cm and 0.3125 cm]
\coordinate[] (v1);
\coordinate[left = of v1] (v0);
\coordinate[right = of v1] (v2);
\coordinate[right = of v2] (v3);
\coordinate[right = of v3] (v8);
\coordinate[below = of v2] (v4);
\coordinate[below = of v4] (v5);
\coordinate[left = of v5] (v6);
\coordinate[left = of v6] (v9);
\coordinate[right = of v5] (v7);
\coordinate[right = of v7] (v10);
\draw[fermionnoarrow] (v0) -- (v8);
\draw[sigma] (v2) -- (v4);
\draw[phi] (v4) -- (v6);
\draw[phi] (v4) -- (v7);
\draw[fermionnoarrow] (v9) -- (v10);
\end{tikzpicture}
}
\parbox{15mm}{
\begin{tikzpicture}[line width=1 pt,node distance=0.5 cm and 0.3125 cm]
\coordinate[] (v1);
\coordinate[left = of v1] (v0);
\coordinate[right = of v1] (v2);
\coordinate[right = of v2] (v3);
\coordinate[right = of v3] (v8);
\coordinate[below = of v2] (v4);
\coordinate[below = of v4] (v5);
\coordinate[left = of v5] (v6);
\coordinate[left = of v6] (v9);
\coordinate[right = of v5] (v7);
\coordinate[right = of v7] (v10);
\draw[fermionnoarrow] (v0) -- (v8);
\draw[phi] (v2) -- (v4);
\draw[sigma] (v4) -- (v6);
\draw[phi] (v4) -- (v7);
\draw[fermionnoarrow] (v9) -- (v10);
\end{tikzpicture}
}\right)
\times
\left(\,\parbox{15mm}{
\begin{tikzpicture}[line width=1 pt,node distance=0.5 cm and 0.3125 cm]
\coordinate[] (v1);
\coordinate[right = of v1] (v2);
\coordinate[right = of v2] (v3);
\coordinate[below = of v2] (v4);
\coordinate[below = of v4] (v5);
\coordinate[left = of v5] (v6);
\coordinate[right = of v5] (v7);
\draw[fermionnoarrow] (v1) -- (v3);
\draw[phi] (v2) -- (v5);
\draw[fermionnoarrow] (v6) -- (v7);
\end{tikzpicture}
}\!\!\!\!\!\!\!\!\!\!\!\!\right)^3\nonumber\\
&&\qquad\qquad=\left\{\ \parbox{15mm}{
\begin{tikzpicture}[line width=1 pt,node distance=0.5 cm and 0.3125 cm]
\coordinate[] (v1);
\coordinate[left = of v1] (v0);
\coordinate[right = of v1] (v2);
\coordinate[right = of v2] (v3);
\coordinate[right = of v3] (v8);
\coordinate[below = of v2] (v4);
\coordinate[below = of v4] (v5);
\coordinate[left = of v5] (v6);
\coordinate[left = of v6] (v9);
\coordinate[right = of v5] (v7);
\coordinate[right = of v7] (v10);
\draw[fermionnoarrow] (v0) -- (v8);
\draw[phi] (v2) -- (v4);
\draw[phi] (v4) -- (v6);
\draw[phi] (v4) -- (v7);
\draw[phi] (v2) -- (v9);
\draw[phi] (v7) -- (v3);
\draw[phi] (v7) -- (v8);
\draw[fermionnoarrow] (v9) -- (v10);
\end{tikzpicture}
}
\parbox{15mm}{
\begin{tikzpicture}[line width=1 pt,node distance=0.5 cm and 0.3125 cm]
\coordinate[] (v1);
\coordinate[left = of v1] (v0);
\coordinate[right = of v1] (v2);
\coordinate[right = of v2] (v3);
\coordinate[right = of v3] (v8);
\coordinate[below = of v2] (v4);
\coordinate[below = of v4] (v5);
\coordinate[left = of v5] (v6);
\coordinate[left = of v6] (v9);
\coordinate[right = of v5] (v7);
\coordinate[right = of v7] (v10);
\draw[fermionnoarrow] (v0) -- (v8);
\draw[phi] (v2) -- (v4);
\draw[phi] (v4) -- (v6);
\draw[phi] (v4) -- (v7);
\draw[phi] (v6) -- (v1);
\draw[phi] (v7) -- (v3);
\draw[phi] (v7) -- (v8);
\draw[fermionnoarrow] (v9) -- (v10);
\end{tikzpicture}
}
\parbox{15mm}{
\begin{tikzpicture}[line width=1 pt,node distance=0.5 cm and 0.3125 cm]
\coordinate[] (v1);
\coordinate[left = of v1] (v0);
\coordinate[right = of v1] (v2);
\coordinate[right = of v2] (v3);
\coordinate[right = of v3] (v8);
\coordinate[below = of v2] (v4);
\coordinate[below = of v4] (v5);
\coordinate[left = of v5] (v6);
\coordinate[left = of v6] (v9);
\coordinate[right = of v5] (v7);
\coordinate[right = of v7] (v10);
\draw[fermionnoarrow] (v0) -- (v8);
\draw[phi] (v2) -- (v4);
\draw[phi] (v4) -- (v6);
\draw[phi] (v4) -- (v7);
\draw[phi] (v6) -- (v1);
\draw[phi] (v7) -- (v3);
\draw[phi] (v3) -- (v10);
\draw[fermionnoarrow] (v9) -- (v10);
\end{tikzpicture}
}
\parbox{15mm}{
\begin{tikzpicture}[line width=1 pt,node distance=0.5 cm and 0.3125 cm]
\coordinate[] (v1);
\coordinate[left = of v1] (v0);
\coordinate[right = of v1] (v2);
\coordinate[right = of v2] (v3);
\coordinate[right = of v3] (v8);
\coordinate[below = of v2] (v4);
\coordinate[below = of v4] (v5);
\coordinate[left = of v5] (v6);
\coordinate[left = of v6] (v9);
\coordinate[right = of v5] (v7);
\coordinate[right = of v7] (v10);
\draw[fermionnoarrow] (v0) -- (v8);
\draw[phi] (v2) -- (v4);
\draw[phi] (v4) -- (v6);
\draw[phi] (v4) -- (v7);
\draw[phi] (v2) -- (v9);
\draw[phi] (v7) -- (v3);
\draw[phi] (v3) -- (v10);
\draw[fermionnoarrow] (v9) -- (v10);
\end{tikzpicture}
}
\parbox{15mm}{
\begin{tikzpicture}[line width=1 pt,node distance=0.5 cm and 0.3125 cm]
\coordinate[] (v1);
\coordinate[left = of v1] (v0);
\coordinate[right = of v1] (v2);
\coordinate[right = of v2] (v3);
\coordinate[right = of v3] (v8);
\coordinate[below = of v2] (v4);
\coordinate[left = of v4] (v11);
\coordinate[below = of v4] (v5);
\coordinate[left = of v5] (v6);
\coordinate[left = of v6] (v9);
\coordinate[right = of v5] (v7);
\coordinate[right = of v7] (v10);
\draw[fermionnoarrow] (v0) -- (v8);
\draw[phi] (v1) -- (v11);
\draw[phi] (v11) -- (v9);
\draw[phi] (v11) -- (v5);
\draw[phi] (v5) -- (v2);
\draw[phi] (v5) -- (v3);
\draw[phi] (v5) -- (v8);
\draw[fermionnoarrow] (v9) -- (v10);
\end{tikzpicture}}
\dots
\parbox{15mm}{
\begin{tikzpicture}[line width=1 pt,node distance=0.5 cm and 0.3125 cm]
\coordinate[] (v1);
\coordinate[left = of v1] (v0);
\coordinate[right = of v1] (v2);
\coordinate[right = of v2] (v3);
\coordinate[right = of v3] (v8);
\coordinate[below = of v2] (v4);
\coordinate[below = of v4] (v5);
\coordinate[left = of v5] (v6);
\coordinate[left = of v6] (v9);
\coordinate[right = of v5] (v7);
\coordinate[right = of v7] (v10);
\draw[fermionnoarrow] (v0) -- (v8);
\draw[phi] (v2) -- (v4);
\draw[A] (v4) -- (v6);
\draw[phi] (v4) -- (v7);
\draw[phi] (v2) -- (v9);
\draw[phi] (v7) -- (v3);
\draw[phi] (v7) -- (v8);
\draw[fermionnoarrow] (v9) -- (v10);
\end{tikzpicture}
}\dots\right\}
\label{eq:tripleN3}
\end{eqnarray}
gives 135 $G^5v^2$ factorisable diagrams\footnote{The dots in the second line in (\ref{eq:tripleN3}) stand for diagrams with the same topology as the first five with the replacement of the bulk triple vertex structure in the first line.}, whose total contribution to the potential is
\begin{align}
V_{N^3G^2v^2}&=\frac{G^5 m_1^5m_2}{r^5}\left[\frac7{12}v^2+\frac53 v_2^2+\frac1{12}{v^r}^2-\frac73{v_2^r}^2\right]\nonumber\\
&+\frac{G^5 m_1^4m_2^2}{r^5}\left[\frac{8 {v^r}^2}{\epsilon}+\frac13\left(13v_1^2-154v_1.v_2-17v_2^2+335{v_1^r}^2-478{v_1^r}{v_2^r} + 277{v_2^r}^2\right)\right]\nonumber\\
&+\frac{G^5 m_1^3m_2^3}{r^5}\!\left[\frac{26 v_1^rv^r}{\epsilon}+\frac12\left(49v_1^2-147v_1.v_2+615{v_1^r}^2-569{v_1^r}{v_2^r} \right)\!\right]+ \!(1\!\leftrightarrow\!2)\,,
\end{align}
with $\frac1\epsilon\equiv \frac1{(d-3)}-5 \log{\left(\frac{r\sqrt{4\pi \gamma_E}}{L}\right)}$, $v_{1,2}^r\equiv\frac{r.v_{1,2}}r$, and $v\equiv v_1-v_2$.

\noindent
The same 7 $G^2v^2$ prime diagrams can also be combined with the 3 static prime $G^3$ diagrams
\begin{align}
\label{eqG2G3}
&\left(\ \parbox{15mm}{
\begin{tikzpicture}[line width=1 pt,node distance=0.5 cm and 0.3125 cm]
\coordinate[] (v1);
\coordinate[left = of v1] (v0);
\coordinate[right = of v1] (v2);
\coordinate[right = of v2] (v3);
\coordinate[right = of v3] (v8);
\coordinate[below = of v2] (v4);
\coordinate[below = of v4] (v5);
\coordinate[left = of v5] (v6);
\coordinate[left = of v6] (v9);
\coordinate[right = of v5] (v7);
\coordinate[right = of v7] (v10);
\draw[fermionnoarrow] (v0) -- (v8);
\draw[phi] (v2) -- (v4);
\draw[phi] (v4) -- (v6);
\draw[phi] (v4) -- (v7);
\draw[fermionnoarrow] (v9) -- (v10);
\end{tikzpicture}
}
\parbox{15mm}{
\begin{tikzpicture}[line width=1 pt,node distance=0.5 cm and 0.3125 cm]
\coordinate[] (v1);
\coordinate[left = of v1] (v0);
\coordinate[right = of v1] (v2);
\coordinate[right = of v2] (v3);
\coordinate[right = of v3] (v8);
\coordinate[below = of v2] (v4);
\coordinate[below = of v4] (v5);
\coordinate[left = of v5] (v6);
\coordinate[left = of v6] (v9);
\coordinate[right = of v5] (v7);
\coordinate[right = of v7] (v10);
\draw[fermionnoarrow] (v0) -- (v8);
\draw[phi] (v2) -- (v4);
\draw[A] (v4) -- (v6);
\draw[phi] (v4) -- (v7);
\draw[fermionnoarrow] (v9) -- (v10);
\end{tikzpicture}
}
\parbox{15mm}{
\begin{tikzpicture}[line width=1 pt,node distance=0.5 cm and 0.3125 cm]
\coordinate[] (v1);
\coordinate[left = of v1] (v0);
\coordinate[right = of v1] (v2);
\coordinate[right = of v2] (v3);
\coordinate[right = of v3] (v8);
\coordinate[below = of v2] (v4);
\coordinate[below = of v4] (v5);
\coordinate[left = of v5] (v6);
\coordinate[left = of v6] (v9);
\coordinate[right = of v5] (v7);
\coordinate[right = of v7] (v10);
\draw[fermionnoarrow] (v0) -- (v8);
\draw[A] (v2) -- (v4);
\draw[phi] (v4) -- (v6);
\draw[phi] (v4) -- (v7);
\draw[fermionnoarrow] (v9) -- (v10);
\end{tikzpicture}
}
\parbox{15mm}{
\begin{tikzpicture}[line width=1 pt,node distance=0.5 cm and 0.3125 cm]
\coordinate[] (v1);
\coordinate[left = of v1] (v0);
\coordinate[right = of v1] (v2);
\coordinate[right = of v2] (v3);
\coordinate[right = of v3] (v8);
\coordinate[below = of v2] (v4);
\coordinate[below = of v4] (v5);
\coordinate[left = of v5] (v6);
\coordinate[left = of v6] (v9);
\coordinate[right = of v5] (v7);
\coordinate[right = of v7] (v10);
\draw[fermionnoarrow] (v0) -- (v8);
\draw[A] (v2) -- (v4);
\draw[A] (v4) -- (v6);
\draw[phi] (v4) -- (v7);
\draw[fermionnoarrow] (v9) -- (v10);
\end{tikzpicture}
}
\parbox{15mm}{
\begin{tikzpicture}[line width=1 pt,node distance=0.5 cm and 0.3125 cm]
\coordinate[] (v1);
\coordinate[left = of v1] (v0);
\coordinate[right = of v1] (v2);
\coordinate[right = of v2] (v3);
\coordinate[right = of v3] (v8);
\coordinate[below = of v2] (v4);
\coordinate[below = of v4] (v5);
\coordinate[left = of v5] (v6);
\coordinate[left = of v6] (v9);
\coordinate[right = of v5] (v7);
\coordinate[right = of v7] (v10);
\draw[fermionnoarrow] (v0) -- (v8);\draw[phi] (v2) -- (v4);
\draw[A] (v4) -- (v6);
\draw[A] (v4) -- (v7);
\draw[fermionnoarrow] (v9) -- (v10);
\end{tikzpicture}
}
\parbox{15mm}{
\begin{tikzpicture}[line width=1 pt,node distance=0.5 cm and 0.3125 cm]
\coordinate[] (v1);
\coordinate[left = of v1] (v0);
\coordinate[right = of v1] (v2);
\coordinate[right = of v2] (v3);
\coordinate[right = of v3] (v8);
\coordinate[below = of v2] (v4);
\coordinate[below = of v4] (v5);
\coordinate[left = of v5] (v6);
\coordinate[left = of v6] (v9);
\coordinate[right = of v5] (v7);
\coordinate[right = of v7] (v10);
\draw[fermionnoarrow] (v0) -- (v8);
\draw[sigma] (v2) -- (v4);
\draw[phi] (v4) -- (v6);
\draw[phi] (v4) -- (v7);
\draw[fermionnoarrow] (v9) -- (v10);
\end{tikzpicture}
}
\parbox{15mm}{
\begin{tikzpicture}[line width=1 pt,node distance=0.5 cm and 0.3125 cm]
\coordinate[] (v1);
\coordinate[left = of v1] (v0);
\coordinate[right = of v1] (v2);
\coordinate[right = of v2] (v3);
\coordinate[right = of v3] (v8);
\coordinate[below = of v2] (v4);
\coordinate[below = of v4] (v5);
\coordinate[left = of v5] (v6);
\coordinate[left = of v6] (v9);
\coordinate[right = of v5] (v7);
\coordinate[right = of v7] (v10);
\draw[fermionnoarrow] (v0) -- (v8);
\draw[phi] (v2) -- (v4);
\draw[sigma] (v4) -- (v6);
\draw[phi] (v4) -- (v7);
\draw[fermionnoarrow] (v9) -- (v10);
\end{tikzpicture}
}\right)
\notag\\
&\qquad\qquad\qquad\qquad\qquad\qquad\qquad\qquad\qquad\qquad\times
\left(\
\parbox{15mm}{
\begin{tikzpicture}[line width=1 pt,node distance=0.5 cm and 0.3125 cm]
\coordinate[] (v1);
\coordinate[right = of v1] (v2);
\coordinate[right = of v2] (v3);
\coordinate[right = of v3] (v4);
\coordinate[right = of v4] (v5);
\coordinate[below = of v1] (v6);
\coordinate[below = of v6] (v7);
\coordinate[right = of v7] (v8);
\coordinate[right = of v8] (v9);
\coordinate[right = of v9] (v10);
\coordinate[right = of v10] (v11);
\coordinate[below = of v2] (v12);
\coordinate[below = of v4] (v13);
\draw[sigma] (v12) -- (v13);
\draw[phi] (v2) -- (v8);
\draw[phi] (v4) -- (v10);
\draw[fermionnoarrow] (v1) -- (v5);
\draw[fermionnoarrow] (v7) -- (v11);
\end{tikzpicture}
}
\parbox{15mm}{
\begin{tikzpicture}[line width=1 pt,node distance=0.333 cm and 0.3125 cm]
\coordinate[] (v1);
\coordinate[right = of v1] (v2);
\coordinate[right = of v2] (v3);
\coordinate[right = of v4] (v5);
\coordinate[below right = of v2] (v6);
\coordinate[below = of v6] (v7);
\coordinate[below left = of v7] (v8);
\coordinate[below right = of v7] (v9);
\coordinate[left = of v8] (v10);
\coordinate[right = of v9] (v11);
\draw[sigma] (v6) -- (v7);
\draw[phi] (v2) -- (v6);
\draw[phi] (v6) -- (v4);
\draw[phi] (v8) -- (v7);
\draw[phi] (v9) -- (v7);
\draw[fermionnoarrow] (v1) -- (v5);
\draw[fermionnoarrow] (v10) -- (v11);
\end{tikzpicture}
} 
\parbox{15mm}{
\begin{tikzpicture}[line width=1 pt,node distance=0.5 cm and 0.25 cm]
\coordinate[] (v1);
\coordinate[right = of v1] (v2);
\coordinate[below = of v2] (v3);
\coordinate[below = of v3] (v4);
\coordinate[right = of v3] (v5);
\coordinate[right = of v5] (v6);
\coordinate[below left = of v6] (v7);
\coordinate[below right = of v6] (v8);
\coordinate[right = of v8] (v9);
\coordinate[left = of v4] (v10);
\coordinate[above right= of v6] (v11);
\coordinate[right= of v11] (v12);
\draw[phi] (v2) -- (v4);
\draw[phi] (v7) -- (v6);
\draw[phi] (v8) -- (v6);
\draw[sigma] (v3) -- (v6);
\draw[fermionnoarrow] (v9) -- (v10);
\draw[fermionnoarrow] (v1) -- (v12);
\end{tikzpicture}
}
\right)
\end{align}
to give 85 $G^5v^2$ factorisable diagrams (17 trivially vanishing because the
central one in the second factor in eq.(\ref{eqG2G3}) is zero at leading order),
whose total contribution to the potential is
\begin{eqnarray}
V_{G^3v^0\ G^2v^2}&=&\frac{G^5 m_1^5m_2}{r^5}\left[\frac76v^2-\frac23v_2^2+\frac16{v^r}^2-\frac23{v_2^r}^2\right]\nonumber\\
&&+\frac{2G^5 m_1^4m_2^2}{3r^5}\left[19v_1^2-70v_1.v_2+7v_2^2+65{v_1^r}^2-82{v_1^r}{v_2^r} +37 {v_2^r}^2\right]\nonumber\\
&&+\frac{G^5 m_1^3m_2^3}{r^5}\left[-3v_1^2-63v_1.v_2+215{v_1^r}^2-153{v_1^r}{v_2^r} \right]
+ (1\!\leftrightarrow\!2)\,.
\end{eqnarray}
\noindent
The combination of two Newtonian diagrams with the 28 $G^3v^2$ prime diagrams
\begin{equation}
\left(
\parbox{18mm}{
$\text{28 $G^3v^2$}$ diagrams}
\right)
\times
\left(\,\parbox{15mm}{
\begin{tikzpicture}[line width=1 pt,node distance=0.5 cm and 0.3125 cm]
\coordinate[] (v1);
\coordinate[right = of v1] (v2);
\coordinate[right = of v2] (v3);
\coordinate[below = of v2] (v4);
\coordinate[below = of v4] (v5);
\coordinate[left = of v5] (v6);
\coordinate[right = of v5] (v7);
\draw[fermionnoarrow] (v1) -- (v3);
\draw[phi] (v2) -- (v5);
\draw[fermionnoarrow] (v6) -- (v7);
\end{tikzpicture}
}\!\!\!\!\!\!\!\!\!\!\!\!\right)^2
\end{equation}
gives 267 $G^5v^2$ factorisable diagrams (11 of which trivially vanishing), whose total contribution to the potential is,
\begin{align}
V_{N^{2}G^{3}v^{2}}&=\frac{G^{5}m_{1}^{5}m_{2}}{r^{5}}\!\bigg[\frac{1}{\epsilon}\!\left(\frac{41}{20}v_{1}^{2}-2v_{1}.v_{2}-\frac{203}{20}{v_{1}^{r}}^{2}+10v_{1}^{r}v_{2}^{r}\right)\nonumber\\
&\qquad\qquad\quad+\left(\frac{3647}{600}v_{1}^{2}-\frac{19}{6}v_{1}.v_{2}-\frac{27971}{600}{v_{1}^{r}}^{2}+\frac{287}{6}v_{1}^{r}v_{2}^{r}\right)\bigg]\nonumber\\
&+\frac{G^{5}m_{1}^{4}m_{2}^{2}}{r^{5}}\bigg[\frac{1}{\epsilon}\left(\frac{287}{10}v_{1}^{2}-28v_{1}.v_{2}-\frac{1191}{10}{v_{1}^{r}}^{2}+97v_{1}^{r}v_{2}^{r}+20{v_{2}^{r}}^{2}\right)\nonumber\\
&\qquad\qquad\quad+\left(\frac{37829}{300}-\frac{3}{2}\pi^{2}\right)v_{1}^{2}+\left(\frac{99}{32}\pi^{2}-\frac{178}{3}\right)v_{1}.v_{2}+\left(41-\frac{3}{2}\pi^{2}\right)v_{2}^{2}\nonumber\\
&\qquad\qquad\quad+\left(\frac{9}{4}\pi^{2}-\frac{242297}{300}\right){v_{1}^{r}}^{2}+\left(\frac{2792}{3}-\frac{129}{32}\pi^{2}\right){v_{1}^{r}}{v_{2}^{r}}+\left(\frac{3}{2}\pi^{2}-167\right){v_{2}^{r}}^{2}\bigg]\nonumber\\
&+\frac{G^{5}m_{1}^{3}m_{2}^{3}}{r^{5}}\bigg[\frac{1}{\epsilon}\left(\frac{369}{20}v_{1}^{2}-18v_{1}.v_{2}-\frac{927}{20}{v_{1}^{r}}^{2}+45v_{1}^{r}v_{2}^{r}\right)\nonumber\\
&\qquad\qquad\quad+\left(\frac{43741}{200}-6\pi^{2}\right)v_{1}^{2}+\left(\frac{99}{16}\pi^{2}-\frac{117}{2}\right)v_{1}.v_{2}\nonumber\\
&\qquad\qquad\quad+\left(\frac{33}{4}\pi^{2}-\frac{227913}{200}\right){v_{1}^{r}}^{2}+\left(\frac{2057}{2}-\frac{141}{16}\pi^{2}\right){v_{1}^{r}}{v_{2}^{r}}\bigg]+(1\!\leftrightarrow\!2)\,.
\end{align}
\noindent
The product of the Newtonian diagram with the 171 $G^4v^2$ prime diagrams
\begin{equation}
\left(
\parbox{15mm}{
171\ $G^4v^2$\\
diagrams}
\right)
\times
\left(\,\parbox{15mm}{
\begin{tikzpicture}[line width=1 pt,node distance=0.5 cm and 0.3125 cm]
\coordinate[] (v1);
\coordinate[right = of v1] (v2);
\coordinate[right = of v2] (v3);
\coordinate[below = of v2] (v4);
\coordinate[below = of v4] (v5);
\coordinate[left = of v5] (v6);
\coordinate[right = of v5] (v7);
\draw[fermionnoarrow] (v1) -- (v3);
\draw[phi] (v2) -- (v5);
\draw[fermionnoarrow] (v6) -- (v7);
\end{tikzpicture}
}\!\!\!\!\!\!\!\!\!\!\!\!\right)
\end{equation}
gives 665 $G^5v^2$ factorisable diagrams (only 4 of which are trivially vanishing), whose total contribution to the potential is
\begin{align}
  \label{eq:NG4v2}
V_{NG^{4}v^{2}}&=\frac{G^{5}m_{1}^{5}m_{2}}{r^{5}}\left[\frac{4}{\epsilon}\left(v_{1}.v-5v_{1}^{r}v^{r}\right)-\frac{2}{3}\left(v_{1}^{2}+5v_{1}.v_{2}-11{v_{1}^{r}}^{2}+17v_{1}^{r}v_{2}^{r}\right)\right]\nonumber\\
&+\frac{G^{5}m_{1}^{4}m_{2}^{2}}{r^{5}}\bigg[\frac{4}{\epsilon}\left(-\frac{16}{15}v_{1}^{2}-\frac{2}{3}v_{1}.v_{2}+2v_{2}^{2}+\frac{47}{5}{v_{1}^{r}}^{2}-\frac{8}{3}{v_{1}^{r}}{v_{2}^{r}}-7{v_{2}^{r}}^{2}\right)\nonumber\\
&\qquad\qquad\quad-\left(\frac{29}{4}\pi^{2}+\frac{538}{75}\right)v_{1}^{2}+\left(\frac{7}{2}\pi^{2}-\frac{184}{3}\right)v_{1}.v_{2}+\left(\frac{11}{4}\pi^{2}-\frac{134}{3}\right)v_{2}^{2}\nonumber\\
&\qquad\qquad\quad+\left(\frac{781}{12}\pi^{2}+\frac{7216}{25}\right){v_{1}^{r}}^{2}\-\left(\frac{451}{6}\pi^{2}+308\right){v_{1}^{r}}{v_{2}^{r}}+\left(\frac{133}{12}\pi^{2}+\frac{212}{3}\right){v_{2}^{r}}^{2}\bigg]\nonumber\\
&+\frac{G^{5}m_{1}^{3}m_{2}^{3}}{r^{5}}\!\bigg[\frac{4}{\epsilon}\!\left(-\frac{34}{15}v_{1}^{2}+5v_{1}.v_{2}+\frac{419}{15}{v_{1}^{r}}^{2}-\frac{92}{3}{v_{1}^{r}}{v_{2}^{r}}\right)\!\nonumber\\
&\qquad\qquad\quad-\left(\frac{7}{12}\pi^{2}+\frac{13012}{75}\right)v_{1}^{2}+\left(\frac{21}{4}\pi^{2}-72\right)\!v_{1}.v_{2}\nonumber\\
&\qquad\qquad\quad+\left(\frac{841}{12}\pi^{2}+\frac{24444}{25}\right){v_{1}^{r}}^{2}-\left(\frac{299}{4}\pi^{2}+\frac{2312}{3}\right){v_{1}^{r}}{v_{2}^{r}}\bigg]+(1\!\leftrightarrow\!2)\,.
\end{align}
The product of the Newtonian diagram, of the diagram with a single $A_i$ propagator, and of the three static prime $G^3$ diagrams
\begin{eqnarray}
\left(\,\parbox{15mm}{
\begin{tikzpicture}[line width=1 pt,node distance=0.5 cm and 0.3125 cm]
\coordinate[] (v1);
\coordinate[right = of v1] (v2);
\coordinate[right = of v2] (v3);
\coordinate[below = of v2] (v4);
\coordinate[below = of v4] (v5);
\coordinate[left = of v5] (v6);
\coordinate[right = of v5] (v7);
\draw[fermionnoarrow] (v1) -- (v3);
\draw[A] (v2) -- (v5);
\draw[fermionnoarrow] (v6) -- (v7);
\end{tikzpicture}
}\!\!\!\!\!\!\!\!\!\!\!\!\right)
\times
\left(\,\parbox{15mm}{
\begin{tikzpicture}[line width=1 pt,node distance=0.5 cm and 0.3125 cm]
\coordinate[] (v1);
\coordinate[right = of v1] (v2);
\coordinate[right = of v2] (v3);
\coordinate[below = of v2] (v4);
\coordinate[below = of v4] (v5);
\coordinate[left = of v5] (v6);
\coordinate[right = of v5] (v7);
\draw[fermionnoarrow] (v1) -- (v3);
\draw[phi] (v2) -- (v5);
\draw[fermionnoarrow] (v6) -- (v7);
\end{tikzpicture}
}\!\!\!\!\!\!\!\!\!\!\!\!\right)
\times
\left(\
\parbox{15mm}{
\begin{tikzpicture}[line width=1 pt,node distance=0.5 cm and 0.3125 cm]
\coordinate[] (v1);
\coordinate[right = of v1] (v2);
\coordinate[right = of v2] (v3);
\coordinate[right = of v3] (v4);
\coordinate[right = of v4] (v5);
\coordinate[below = of v1] (v6);
\coordinate[below = of v6] (v7);
\coordinate[right = of v7] (v8);
\coordinate[right = of v8] (v9);
\coordinate[right = of v9] (v10);
\coordinate[right = of v10] (v11);
\coordinate[below = of v2] (v12);
\coordinate[below = of v4] (v13);
\draw[sigma] (v12) -- (v13);
\draw[phi] (v2) -- (v8);
\draw[phi] (v4) -- (v10);
\draw[fermionnoarrow] (v1) -- (v5);
\draw[fermionnoarrow] (v7) -- (v11);
\end{tikzpicture}
}
\parbox{15mm}{
\begin{tikzpicture}[line width=1 pt,node distance=0.333 cm and 0.3125 cm]
\coordinate[] (v1);
\coordinate[right = of v1] (v2);
\coordinate[right = of v2] (v3);
\coordinate[right = of v4] (v5);
\coordinate[below right = of v2] (v6);
\coordinate[below = of v6] (v7);
\coordinate[below left = of v7] (v8);
\coordinate[below right = of v7] (v9);
\coordinate[left = of v8] (v10);
\coordinate[right = of v9] (v11);
\draw[sigma] (v6) -- (v7);
\draw[phi] (v2) -- (v6);
\draw[phi] (v4) -- (v6);
\draw[phi] (v8) -- (v7);
\draw[phi] (v9) -- (v7);
\draw[fermionnoarrow] (v1) -- (v5);
\draw[fermionnoarrow] (v10) -- (v11);
\end{tikzpicture}
}
\parbox{15mm}{
\begin{tikzpicture}[line width=1 pt,node distance=0.5 cm and 0.25 cm]
\coordinate[] (v1);
\coordinate[right = of v1] (v2);
\coordinate[below = of v2] (v3);
\coordinate[below = of v3] (v4);
\coordinate[right = of v3] (v5);
\coordinate[right = of v5] (v6);
\coordinate[below left = of v6] (v7);
\coordinate[below right = of v6] (v8);
\coordinate[right = of v8] (v9);
\coordinate[left = of v4] (v10);
\coordinate[above right= of v6] (v11);
\coordinate[right= of v11] (v12);
\draw[phi] (v2) -- (v4);
\draw[phi] (v7) -- (v6);
\draw[phi] (v8) -- (v6);
\draw[sigma] (v3) -- (v6);
\draw[fermionnoarrow] (v9) -- (v10);
\draw[fermionnoarrow] (v1) -- (v12);
\end{tikzpicture}
} \right)
\end{eqnarray}
gives 27 factorisable diagrams (5 of which are trivially vanishing), whose total contribution to the potential is
\begin{align}
V_{NA\,G^3v^0}&=\frac43\frac{G^5 m_1^5m_2}{r^5}v_1.v_2+\frac{152}3\frac{G^5 m_1^4m_2^2}{r^5}v_1.v_2+76\frac{G^5 m_1^3m_2^3}{r^5}v_1.v_2 +
(1\!\leftrightarrow\!2)\,.
\end{align}
\noindent
Finally, there are 25 diagrams which allow a non-vanishing static limit, thus contributing also at the 4PN dynamics (as computed in \cite{Foffa:2016rgu}).
The static limit being just the leading order term of an expansion in the parameter $v^2$ (see the Appendix for more details), the next-to-leading order part of such diagrams contributes to the 5PN sector, and its evaluation can also be carried along the factorisation technique used so far.
Hence, the 19 4PN diagrams resulting from the product of the three $G^3v^0$ with two Newtonian
\begin{eqnarray}
\left(\,\parbox{15mm}{
\begin{tikzpicture}[line width=1 pt,node distance=0.5 cm and 0.3125 cm]
\coordinate[] (v1);
\coordinate[right = of v1] (v2);
\coordinate[right = of v2] (v3);
\coordinate[below = of v2] (v4);
\coordinate[below = of v4] (v5);
\coordinate[left = of v5] (v6);
\coordinate[right = of v5] (v7);
\draw[fermionnoarrow] (v1) -- (v3);
\draw[phi] (v2) -- (v5);
\draw[fermionnoarrow] (v6) -- (v7);
\end{tikzpicture}
}\!\!\!\!\!\!\!\!\!\!\!\!\right)^2
\times
\left(\
\parbox{15mm}{
\begin{tikzpicture}[line width=1 pt,node distance=0.5 cm and 0.3125 cm]
\coordinate[] (v1);
\coordinate[right = of v1] (v2);
\coordinate[right = of v2] (v3);
\coordinate[right = of v3] (v4);
\coordinate[right = of v4] (v5);
\coordinate[below = of v1] (v6);
\coordinate[below = of v6] (v7);
\coordinate[right = of v7] (v8);
\coordinate[right = of v8] (v9);
\coordinate[right = of v9] (v10);
\coordinate[right = of v10] (v11);
\coordinate[below = of v2] (v12);
\coordinate[below = of v4] (v13);
\draw[sigma] (v12) -- (v13);
\draw[phi] (v2) -- (v8);
\draw[phi] (v4) -- (v10);
\draw[fermionnoarrow] (v1) -- (v5);
\draw[fermionnoarrow] (v7) -- (v11);
\end{tikzpicture}
}
\parbox{15mm}{
\begin{tikzpicture}[line width=1 pt,node distance=0.333 cm and 0.3125 cm]
\coordinate[] (v1);
\coordinate[right = of v1] (v2);
\coordinate[right = of v2] (v3);
\coordinate[right = of v4] (v5);
\coordinate[below right = of v2] (v6);
\coordinate[below = of v6] (v7);
\coordinate[below left = of v7] (v8);
\coordinate[below right = of v7] (v9);
\coordinate[left = of v8] (v10);
\coordinate[right = of v9] (v11);
\draw[sigma] (v6) -- (v7);
\draw[phi] (v2) -- (v6);
\draw[phi] (v4) -- (v6);
\draw[phi] (v8) -- (v7);
\draw[phi] (v9) -- (v7);
\draw[fermionnoarrow] (v1) -- (v5);
\draw[fermionnoarrow] (v10) -- (v11);
\end{tikzpicture}
}
\parbox{15mm}{
\begin{tikzpicture}[line width=1 pt,node distance=0.5 cm and 0.25 cm]
\coordinate[] (v1);
\coordinate[right = of v1] (v2);
\coordinate[below = of v2] (v3);
\coordinate[below = of v3] (v4);
\coordinate[right = of v3] (v5);
\coordinate[right = of v5] (v6);
\coordinate[below left = of v6] (v7);
\coordinate[below right = of v6] (v8);
\coordinate[right = of v8] (v9);
\coordinate[left = of v4] (v10);
\coordinate[above right= of v6] (v11);
\coordinate[right= of v11] (v12);
\draw[phi] (v2) -- (v4);
\draw[phi] (v7) -- (v6);
\draw[phi] (v8) -- (v6);
\draw[sigma] (v3) -- (v6);
\draw[fermionnoarrow] (v9) -- (v10);
\draw[fermionnoarrow] (v1) -- (v12);
\end{tikzpicture}
} \right)
\end{eqnarray}
give, at next-to leading order in $v^2$
\begin{align}
V_{N^{2}G^{3}v^{0}}^{NLO}&=\frac{G^{5}m_{1}^{5}m_{2}}{r^{5}}\!\bigg[\frac{1}{6\epsilon}\!\left(\!v_{1}.v_{2}-\frac{13}{10}v_{1}^{2}+\frac{59}{10}{v_{1}^{r}}^{2}\!-5v_{1}^{r}v_{2}^{r}\!\right)\!-\frac{3691}{1800}v_{1}^{2}+\frac{17}{36}v_{1}.v_{2}\notag\\
&\qquad\qquad\quad-\frac{9}{4}v_{2}^{2}+\frac{6113}{1800}{v_{1}^{r}}^{2}\!-\frac{115}{36}v_{1}^{r}v_{2}^{r}\!\bigg]\nonumber\\&
+\frac{G^{5}m_{1}^{4}m_{2}^{2}}{r^{5}}\!\bigg[\frac{1}{3\epsilon}\!\left(7v_{1}.v_{2}-\frac{91}{10}v_{1}^{2}+\frac{323}{10}{v_{1}^{r}}^{2}\!-26{v_{1}^{r}}{v_{2}^{r}}\right)\notag\\
&\qquad\qquad\quad+\left(\frac{7}{32}\pi^{2}\!-\frac{12329}{300}\right)\!v_{1}^{2}+\left(\!\frac{97}{18}-\frac{17}{32}\pi^{2}\right)\!v_{1}.v_{2}\nonumber\\
&\qquad\qquad\quad+\left(\frac{7}{32}\pi^{2}-\!\frac{98}{9}\right)v_{2}^{2}+\left(\frac{26891}{900}-\!\frac{39}{32}\pi^{2}\right)\!{v_{1}^{r}}^{2}\notag\\
&\qquad\qquad\quad+\left(\frac{63}{32}\pi^{2}-\!\frac{409}{18}\right)\!{v_{1}^{r}}{v_{2}^{r}}+\left(\frac{8}{3}-\!\frac{15}{32}\pi^{2}\right)\!{v_{2}^{r}}^{2}\!\bigg]\nonumber\\
&+\frac{G^{5}m_{1}^{3}m_{2}^{3}}{r^{5}}\bigg[\frac{1}{\epsilon}\left(-\frac{39}{20}v_{1}^{2}+\frac{3}{2}v_{1}.v_{2}+\frac{191}{60}{v_{1}^{r}}^{2}-\frac{11}{6}v_{1}^{r}v_{2}^{r}\right)+\left(\frac{7}{8}\pi^{2}-\frac{25669}{1800}\right)v_{1}^{2}\nonumber\\
&\qquad\qquad\quad+\left(-\frac{17}{16}\pi^{2}+\frac{65}{36}\right)v_{1}.v_{2}+\left(-\frac{53}{16}\pi^{2}+\frac{1713}{200}\right){v_{1}^{r}}^{2}\nonumber\\&\qquad\qquad\quad+\left(\frac{31}{8}\pi^{2}+\frac{29}{4}\right){v_{1}^{r}}{v_{2}^{r}}\bigg]+(1\!\leftrightarrow\!2)\,.
\end{align}
To conclude, the 6 4PN diagrams resulting from the combination of five Newtonian diagrams
\begin{eqnarray}
\left(\,\parbox{15mm}{
\begin{tikzpicture}[line width=1 pt,node distance=0.5 cm and 0.3125 cm]
\coordinate[] (v1);
\coordinate[right = of v1] (v2);
\coordinate[right = of v2] (v3);
\coordinate[below = of v2] (v4);
\coordinate[below = of v4] (v5);
\coordinate[left = of v5] (v6);
\coordinate[right = of v5] (v7);
\draw[fermionnoarrow] (v1) -- (v3);
\draw[phi] (v2) -- (v5);
\draw[fermionnoarrow] (v6) -- (v7);
\end{tikzpicture}
}\!\!\!\!\!\!\!\!\!\!\!\!\right)^5
\end{eqnarray}
\noindent 
give, still at next-to-leading order in $v^2$
\begin{align}
V_{N^{6}}^{NLO}&=\frac{G^{5}m_{1}^{5}m_{2}}{r^{5}}\left[-\frac{1}{16}v_{1}^{2}-\frac{1}{48}v_{1}.v_{2}-\frac{81}{80}v_{2}^{2}-\frac{1}{12}{v_{1}^{r}}^{2}+\frac{5}{48}v_{1}^{r}v_{2}^{r}\right]\nonumber\\
&+\frac{G^{5}m_{1}^{4}m_{2}^{2}}{r^{5}}\left[-\frac{5}{3}v_{1}.v_{2}+2v_{2}^{2}-3{v_{1}^{r}}^{2}+\frac{13}{3}{v_{1}^{r}}{v_{2}^{r}}+\frac{1}{3}{v_{2}^{r}}^{2}\right]\nonumber\\
&+\frac{G^{5}m_{1}^{3}m_{2}^{3}}{r^{5}}\left[-\frac{9}{8}v_{1}^{2}-\frac{45}{16}v_{1}.v_{2}-\frac{11}{4}{v_{1}^{r}}^{2}+\frac{89}{16}{v_{1}^{r}}{v_{2}^{r}}\right]+(1\!\leftrightarrow\!2)\,.
\end{align}
Summing all the above contributions, we obtain the total contributions of factorisable diagrams to the $G^5 v^2$ sector of the 5PN potential
\begin{align}
V_{v^{2},\text{fact}}^{5PN}&=\frac{G^{5}m_{1}^{5}m_{2}}{r^{5}}\bigg[\frac{35}{6\epsilon}\left(v_{1}.v-5v_{1}^{r}v^{r}\right)+\frac{727}{144}v_{1}^{2}-\frac{1159}{144}v_{1}.v_{2}-\frac{41}{80}v_{2}^{2}\nonumber\\
&-\frac{643}{18}{v_{1}^{r}}^{2}+\frac{4739}{144}v_{1}^{r}v_{2}^{r}-\frac{11}{4}{v_{2}^{r}}^{2}\bigg]\nonumber\\&+\frac{G^{5}m_{1}^{4}m_{2}^{2}}{r^{5}}\bigg[\frac{1}{\epsilon}\left(\frac{107}{5}v_{1}^{2}-\frac{85}{3}v_{1}.v_{2}+8v_{2}^{2}-\frac{941}{15}{v_{1}^{r}}^{2}+\frac{185}{3}{v_{1}^{r}}{v_{2}^{r}}\right)\nonumber\\
&\qquad\qquad\quad+\left(-\frac{273}{32}\pi^{2}+\frac{7112}{75}\right)v_{1}^{2}+\left(\frac{97}{16}\pi^{2}-\frac{2717}{18}\right)v_{1}.v_{2}\nonumber\\
&\qquad\qquad\quad+\left(\frac{47}{32}\pi^{2}-\frac{122}{9}\right)v_{2}^{2}+\left(\frac{6347}{96}\pi^{2}-\frac{75856}{225}\right){v_{1}^{r}}^{2}\nonumber\\
&\qquad\qquad\quad+\left(-\frac{3707}{48}\pi^{2}+\frac{7025}{18}\right){v_{1}^{r}}{v_{2}^{r}}+\left(\frac{1163}{96}\pi^{2}+\frac{71}{3}\right){v_{2}^{r}}^{2}\bigg]\nonumber\\&+\frac{G^{5}m_{1}^{3}m_{2}^{3}}{r^{5}}\bigg[\frac{1}{2\epsilon}\left(\frac{223}{15}v_{1}^{2}+7v_{1}.v_{2}+\frac{2837}{15}{v_{1}^{r}}^{2}-211{v_{1}^{r}}{v_{2}^{r}}\right)\nonumber\\
&\qquad\qquad\quad+\left(\frac{92387}{1800}-\frac{137}{24}\pi^{2}\right)v_{1}^{2}+\left(\frac{83}{8}\pi^{2}-\frac{24409}{144}\right)v_{1}.v_{2}\nonumber\\
&\qquad\qquad\quad+\left(\frac{3601}{48}\pi^{2}+\frac{36651}{100}\right){v_{1}^{r}}^{2}-\left(\frac{1275}{16}\pi^{2}+\frac{8009}{48}\right){v_{1}^{r}}{v_{2}^{r}}\bigg]\nonumber\\&\qquad\qquad\quad+(1\leftrightarrow2)\,.
\end{align}

\section{Conclusion}
\label{sec:conclusion}

In the recent years, we have observed the emergence of novel ideas 
to provide precision calculations of the two-body dynamics at increasingly
higher PN orders, even though to bootstrap the available techniques one needs to overcome 
three kinds of obstacles. 

The first two main kinds of obstacle standing on the way of scaling precision computations of the two-body dynamics to higher order are
intrinsically difficult master integrals (either in coordinate or in momentum space), 
and proliferation of terms.
The former affect in equal measure all different methodologies employed so far, 
and the next hard step will be met in the high $G$ orders sectors at 6PN.

\begin{figure}[h]
\centering
\includegraphics[width=.3\linewidth]{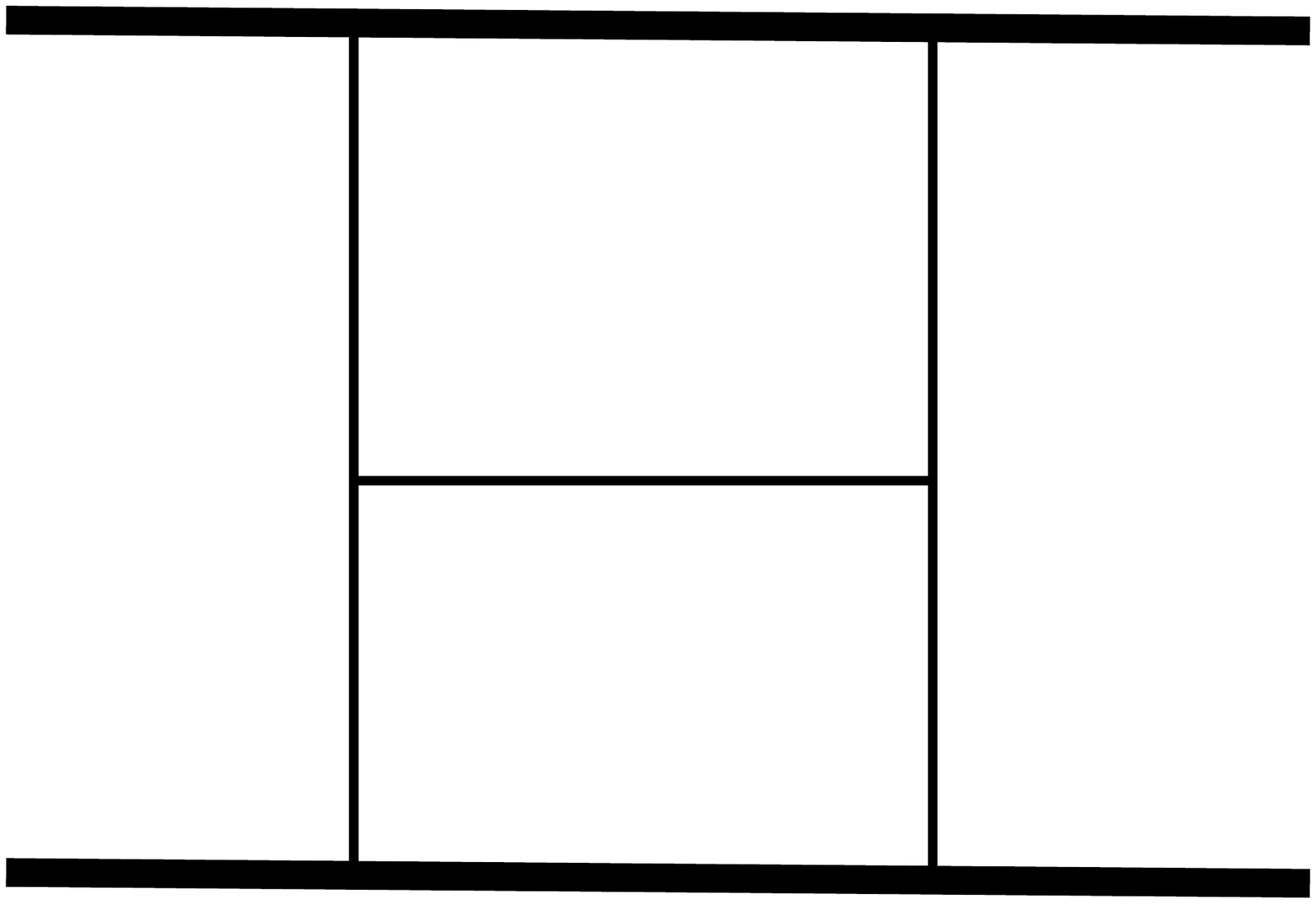}
\includegraphics[width=.3\linewidth]{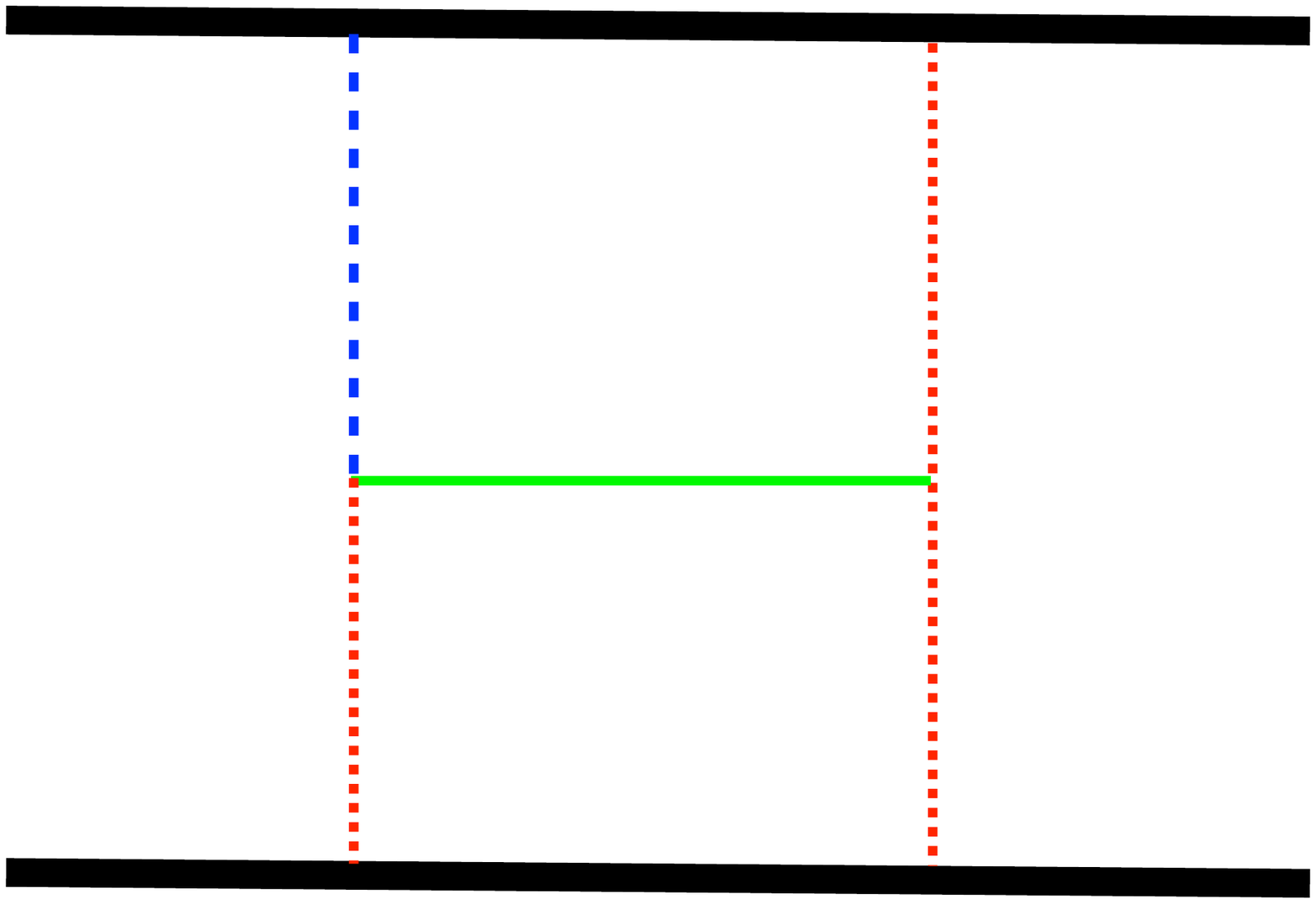}
\caption{On the right, one of the hundreds of possible diagrams deriving from the 3PM topology on the left.}
\label{topgra}
\end{figure}

On the contrary,  the latter impacts unequally different observables and approaches.
For instance, the EFT determination of Lagrangian effective potential involves a 
very large number of diagrams compared to other techniques, as 
gauge freedom, while providing several useful consistency checks,
disperses information among several non-essential contributions.
Furthermore, as noted throughout this paper, 
the explicit velocity counting, usually employed in the PN version of the EFT approach, 
involves a major proliferation of diagrams,
only partially tampered by an efficient metric parameterisation, with respect to a full PM calculation.
This is because several PN diagrams correspond to the same PM ones
(also called topologies in our previous works), as depicted in figure~\ref{topgra}
and this is the price to pay for dealing with simpler integrals with respect to full PM calculations. 
In effect, a reasonable price so far, 
considered the success of velocity-truncation approaches up to 4PN.

The third obstacle is rooted in the observable itself, the binding energy which, as mentioned in the introduction, appears to be less efficient than the scattering angle in storing information.
Therefore, several contributions at a given PN or PM order are actually simple algebraic functions of lower order ones, meaning that a good share of the diagram (or topology) proliferation could actually be overcome by more efficient bookkeeping algorithms.
In view of this observation and former studies, 
we have explicitly exposed this mechanism by reducing the computation of 
all factorisable diagrams as algebraic functions of lower order ones. 
We have started with the static limit~\cite{Foffa:2019hrb}
and, in the present work, we have extended our formalism to the dynamic limit as well,
thus paving the way to provide PN calculations in full generality. 

The ratio of factorisable over non-factorisable diagrams or topologies is a generally growing function of the PN or PM order.  The factorisable vs. non-factorisable diagram number ratio goes from $0.3$ to $0.6$ passing from 2PN to 4PN~\footnote{The comparison must be made between PN or PM orders with the same parity.}; considering diagrams with two powers of velocity, like the one computed in this paper, the ratio grows from $0.6$ to $1.6$ from 3PN to 5PN, while in the static case the progression is from $2.75$ to $6.5$, thus suggesting a growing trend of this ratio, implying a higher and higher relative weight of factorisable diagrams with respect to non-factorisable ones, at higher powers of $G$.
Counting the number of topologies instead of diagrams,  there are 118 factorisable topologies contributing at 5PN, vs. 62 non-factorisable ones.
These numbers indicate that any strategy or approach to scale perturbative calculations at higher orders will benefit from optimisation procedures like the one put forward in this work.

The work~\cite{Blumlein:2020pyo} appeared online on the arXiv on the same day of
this paper, containing the result of the full near zone computation of the
fifth post-Newtonian order two-body dynamics.
We would like to point out that the results presented in this paper,
corresponding to the factorisable diagrams, have been 
explicitly verified via private communications with the authors of~\cite{Blumlein:2020pyo},
finding  complete agreement with their corresponding subset of diagrams.

\acknowledgments
We wish to thank Pierpaolo Mastrolia and Christian Sturm for countless discussions and unwavering support during the completion of this work.
We thank the authors of \cite{Blumlein:2020pyo} for helping us correct eq.~(\ref{eq:NG4v2}) in the first arXiv
version of this paper.
S.F. is supported by the Fonds National Suisse and by the SwissMap NCCR.
R.S. is partially supported by CNPq.
W.J.T. has been supported in part by Grants No. FPA2017-84445-P, SEV-2014-0398 (AEI/ERDF, EU),
the COST Action CA16201 PARTICLEFACE and the ``Juan de la Cierva Formaci\'on'' program (FJCI-2017-32128).

\appendix
\section{Detailed computation of a sample diagram}
\label{appendix}
\begin{figure}[htb]
\centering
\includegraphics[scale=0.18]{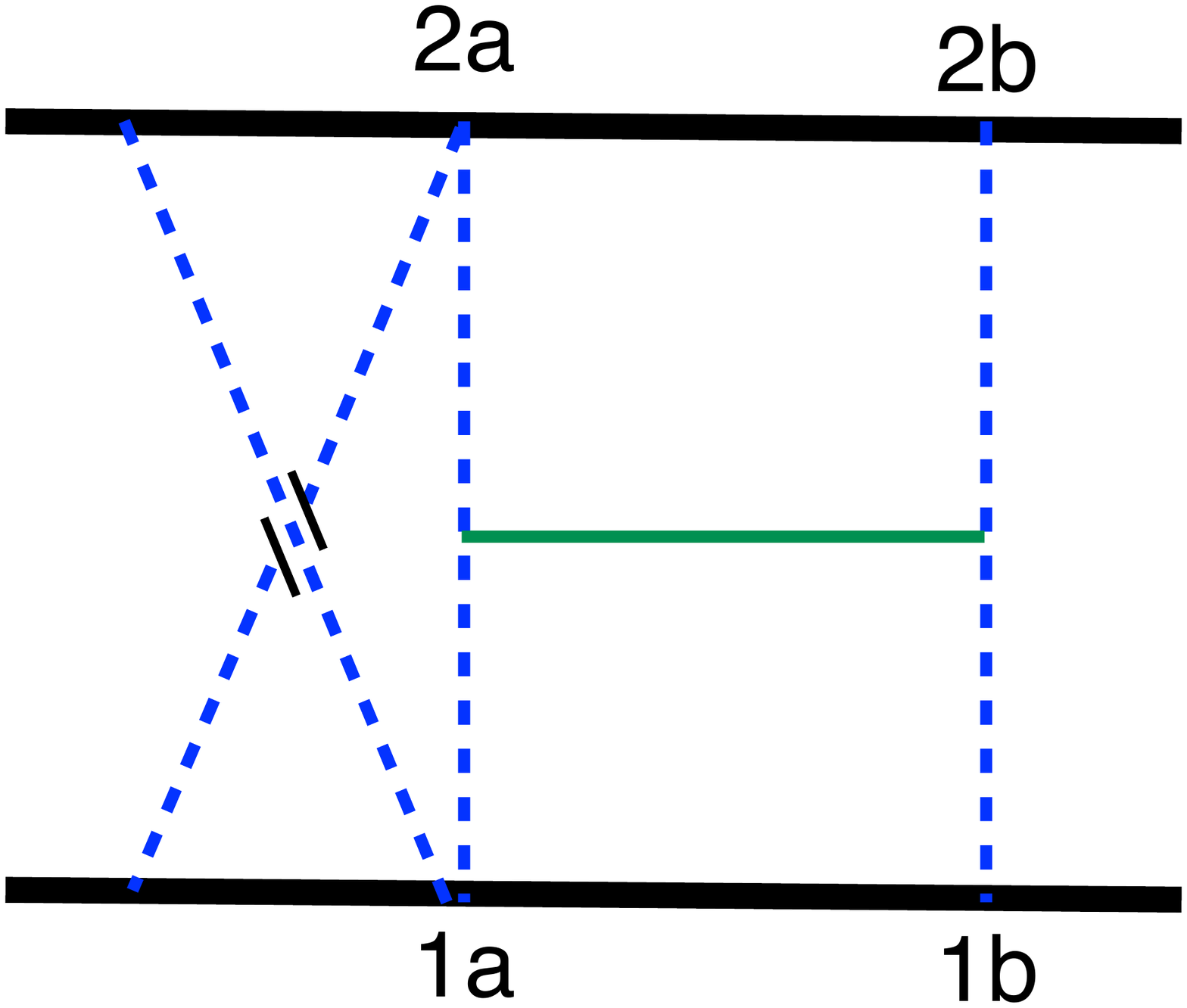}
\caption{Representative diagram at 4PN.}
\label{fig:figapp}
\end{figure}

In this appendix, we explicitly show how the procedure described in section~\ref{sec:procedure} 
works in detail on the diagram of figure~\ref{fig:figapp}.
\par\noindent
The diagram of figure~\ref{fig:figapp} is the product of an H-shaped diagram 
with two Newtonian ones (one attached at the upper left vertex, another attached at the lower left one).
This belong to the most difficult category, because  all factors have to be evaluated at next-to-leading order in time derivatives, meaning that derivative operators
act in every direction (from the H-shaped to the Newtonians, and vice-versa). 
The majority of the other diagrams have derivative operators, when present at all,
acting only in one direction, from the higher-loop factor towards the Newtonians.

We can write each of the Newtonian building blocks as
\be
N=N_0+N_{v^2}\,,
\ee
where
\be
N_0&=&-\frac{G m_1 m_2}r\left[1-(d-3)\left(\log{r}-\frac12\right)\right]\,,\\
N_{v^2}&=&-\frac{G m_1 m_2}{2r}\left[3v_1^2+3v_2^2+v_1.v_2-v_1^r v_2^r- r D_2 v_1^r+r D_1 v_2 -r^2 D_1 D_2\right]\,,
\ee
and we are setting here $L=\sqrt{4\pi \gamma_E}$ in order to simplify notations.
The $N_{v^2}$ term, that is the next-to-leading term of the Newtonian diagram in the velocity expansion, comes from the expansion of the propagator around $k_0\sim 0$, e.g. for the $\phi$ propagator $P_\phi$:
\be
\ba{rcl}
P_{\phi}[t-t',x(t)-x'(t')]&=&\ds -\frac 1{2c_d}\int \frac{d\omega}{2\pi} \frac{d^{\rm d}k}{(2\pi)^{\rm d}}\frac{e^{ik_0(t-t')-ik\cdot [x(t)-x'(t')]}}{k^2-k_0^2}\\
&\simeq&\ds \int \frac{d^{\rm d}k}{\pa{2\pi}^{\rm d}}
\frac{e^{-ik\cdot [x(t)-x'(t')]}}{k^2}\pa{1+\frac{\partial_{t}\partial_{t'}}{k^2}+\ldots}\delta(t-t')\,,
\ea
\ee
which generate velocity terms when the time derivatives act on the particle worldlines present in the exponential factor.

We will also need the derivative of the static term (the derivative of $N_{v^2}$ would obviously be $\mathcal{O}(v^2)\times N_{v^2}$, therefore it is not needed here):
\be
\dot{N}_0=\frac{G m_1 m_2}{r^2}v^r\left[1-(d-3)\left(\log{r}-\frac32\right)\right]\,.
\ee
As to the H-shaped diagram, it reads
\be
F_3=\frac{G^3 m_1^2 m_2^2}{r^3}\left[f_0+f_{v^2} + r f_i D_i +r^2 f_{ij} D_i D_j\right]\,,
\ee
with $i,j=1a,2a,1b,2b$ and $f_0=-1$.
As to $f_{v^2}, f_i, f_{ij}$, they are velocity-dependent terms generated by the same propagator expansion described above,
as well as by a time dependent part contained in the $\phi^2\sigma$ bulk interaction vertex.
They read
\be
&& f_{v^2}=-\frac{161}{36}\left(v_1^2+v_2^2\right)+\frac{25}{36}v_1.v_2 +\frac{29}{12}\left({v_1^r}^2+{v_2^r}^2\right)-\frac{49}{12}v_1^r v_2^r
+\frac{\pi^2}{16}\left(v^2-3{v^r}^2\right)\,,\nonumber\\
&& f_{1a}= \left(\frac49 v_1^r+\frac{11}{36}v_2^r-\frac{\pi^2}{16}v_2^r\right)\,,\nonumber\\
&& f_{2a}=- \left(\frac49 v_2^r+\frac{11}{36}v_1^r-\frac{\pi^2}{16}v_1^r\right)\,,\nonumber\\
&& f_{1a,2a}=\left(-\frac1{3(d-3)}+\log{r}-\frac{29}{18}+\frac{\pi^2}8\right)\,,
\ee
while we can ignore the other $f_i$'s because the corresponding operators do not act on anything.
For the same reason, we can also set
$D_1=0$ in the Newtonian sub-diagram attached to the upper vertex, and
analogously $D_2=0$ for the other one.
We also observe that derivative operators of each factor act just on the static
part of the other factors (again, because we are working at ${\cal O}(v^2)$)
so, for the sake of computing the generalised product $N\times N\times F_3$, we
can make the following replacements:
\begin{subequations}
\be
&&D_1[{\rm lower\ Newtonian\ factor}]\rightarrow -4\frac{v_r}r\,,\\
&&D_2[{\rm upper\ Newtonian\ factor}]\rightarrow -4\frac{v_r}r\,,
\ee
and
\be
D_{1a}\,,D_{2a}[{\rm H\ diagram}]\rightarrow \frac{\dot{N_0}}{N_0}\,.
\ee
\end{subequations}
This brings to
\begin{align}
N\times N\times F_{3}&=\frac{G^{3}m_{1}^{2}m_{2}^{2}}{r^{3}}\left(N_{0}+\left.N_{v^{2}}\right|_{D_1=-4\frac{v^{r}}{r}\,,D_2=0}\right)\left(N_{0}+\left.N_{v^{2}}\right|_{D_2=-4\frac{v^{r}}{r}\,,D_1=0}\right)\notag\\&\qquad\qquad\times\left[-1+f_{v^{2}}+rf_{1a}\frac{\dot{N_{0}}}{N_{0}}+rf_{2a}\frac{\dot{N_{0}}}{N_{0}}+r^{2}f_{1a,2a}\left(\frac{\dot{N_{0}}}{N_{0}}\right)^{2}\right]\notag\\&\simeq\frac{G^{3}m_{1}^{2}m_{2}^{2}}{r^{3}}\Bigg[N_{0}^{2}\left(-1+f_{v^{2}}\right)+N_{0}\dot{N_{0}}\left(rf_{1a}+rf_{2a}\right)\notag\\&\qquad\qquad-N_{0}\left(\left.N_{v^{2}}\right|_{D_2=0}^{D_1=-4\frac{v^{r}}{r}}+\left.N_{v^{2}}\right|_{D_2=-4\frac{v^{r}}{r}}^{D_1=0}\right)+r^{2}f_{1a,2a}\left(\dot{N_{0}}\right)^{2}\Bigg]\,.
\end{align}
Since the only pole is in $f_{1a,2a}$, we can trade $N_0$  and its derivative for their three dimensional value everywhere {\it except} in the last term, where we have to keep the full
${\cal O}(d-3)$ expressions,

\begin{align}
N\times N\times F_{3}\simeq&\frac{G^{5}m_{1}^{4}m_{2}^{4}}{r^{5}}\left\{-1+f_{v^{2}}-\frac{v^{r}}{r}\left(rf_{1a}+rf_{2a}\right)\right.\nonumber\\&\qquad\qquad-\left(3v_{1}^{2}+3v_{2}^{2}+v_{1}.v_{2}-v_{1}^{r}v_{2}^{r}+2v_{1}^{r}v^{r}-2v_{2}^{r}v^{r}\right)\nonumber\\&\qquad\qquad\left.+r^{2}f_{1a,2a}\left(\frac{v^{r}}{r}\right)^{2}\left[1-2(d-3)\left(\log{r}-\frac{3}{2}\right)\right]\right\}\nonumber\\&\simeq\frac{G^{5}m_{1}^{4}m_{2}^{4}}{r^{5}}\Big\{-1+f_{v^{2}}-\left(\frac{5}{36}+\frac{\pi^{2}}{16}\right){v^{r}}^{2}\nonumber\\&\qquad\qquad-\left(3v_{1}^{2}+3v_{2}^{2}+v_{1}.v_{2}+2{v_{1}^{r}}^{2}+2{v_{2}^{r}}^{2}-5v_{1}^{r}v_{2}^{r}\right)\nonumber\\&\qquad\qquad\left.+{v^{r}}^{2}\Bigg[-\frac{1}{3(d-3)}+\log{r}-\frac{29}{18}+\frac{\pi^{2}}{8}+\frac{2}{3}\left(\log{r}-\frac{3}{2}\right)+{\cal O}(\epsilon)\Bigg]\right\}\nonumber\\&\simeq\frac{G^{5}m_{1}^{4}m_{2}^{4}}{r^{5}}\Bigg[-\frac{{v^{r}}^{2}}{3\epsilon}-1-\frac{269}{36}\left(v_{1}^{2}+v_{2}^{2}\right)-\frac{11}{36}v_{1}.v_{2}\nonumber\\&\qquad\qquad-\frac{7}{3}\left({v_{1}^{r}}^{2}+{v_{2}^{r}}^{2}\right)+\frac{77}{12}v_{1}^{r}v_{2}^{r}+\frac{\pi^{2}}{16}\left(v^{2}-2{v^{r}}^{2}\right)\Bigg]\,.
\end{align}
Next, we have to multiply by the symmetry factor and by the vertex corrections
up to $O(v^2)$:

\begin{align}
C^{\rm factorisable}\times\left(\frac{V_{\phi^2}}{V_{\phi}^2}\right)_1\times\left(\frac{V_{\phi^2}}{V_{\phi}^2}\right)_2&=8\frac{1-\frac{d^2}{2(d-2)^2} v_1^2}{2 m_1\left(1+\frac{d}{2(d-2)}v_1^2\right)^2}
\frac{1-\frac{d^2}{2(d-2)^2} v_2^2}{2 m_2\left(1+\frac{d}{2(d-2)}v_2^2\right)^2}\nonumber\\
&\simeq \frac2{m_1 m_2}\left[1-\frac{3d(2d-4)}{2(d-2)^2}\left(v_1^2 +v_2^2\right)\right]\notag\\
&=\frac2{m_1 m_2}\left[1-\frac{15}2\left(v_1^2 +v_2^2\right)+{\cal O}(d-3)\right]\,,
\end{align}
and we can neglect the ${\cal O}(d-3)$ part because the pole in the previous expression is already ${\cal O}(v^2)$.
Hence, the final result gives
\be
\ba{rcl}
N\times N\times F_3&=&\ds\frac{2G^{5}m_{1}^{3}m_{2}^{3}}{r^{5}}\left[\!-\frac{{v^{r}}^{2}}{3\epsilon}\!-1+\frac{1}{36}\left(v_{1}^{2}+v_{2}^{2}\right)\!-\frac{11}{36}v_{1}.v_{2}\right.\\
  &&\ds\qquad\left.-\frac{7}{3}\!\left({v_{1}^{r}}^{2}+{v_{2}^{r}}^{2}\right)
+\frac{77}{12}v_{1}^{r}v_{2}^{r}+\frac{\pi^{2}}{16}\!\pa{v^{2}-2{v^{r}}^{2}}\!\right]\,.
\ea
\ee
The operations described in this appendix were straightforwardly automated 
and adapted to all the 1220 factorisable diagram (the case displayed here is
the most involved one), thus reducing the evaluation of all of them to a few
minutes run. 

\bibliographystyle{JHEP}

\providecommand{\href}[2]{#2}\begingroup\raggedright\endgroup

\end{document}